\documentclass[draftcls,12pt,onecolumn]{IEEEtran}
\ifCLASSINFOpdf
\usepackage{amsmath}
\usepackage{amssymb}
\usepackage{color,colortbl}
\usepackage{cite}
\usepackage{soul}
\usepackage{subcaption}
\usepackage{pgfplots}
\usepackage{tikz}
\usetikzlibrary{calc}
\usepackage{multirow}
\usepackage{booktabs}

\usepackage{tikz}
\usepackage{environ}
\makeatletter
\newsavebox{\measure@tikzpicture}
\NewEnviron{scaletikzpicturetowidth}[1]{%
  \def\tikz@width{#1}%
  \begin{lrbox}{\measure@tikzpicture}%
  \BODY
  \end{lrbox}%
  \pgfmathparse{#1/\wd\measure@tikzpicture}%
  \BODY
}
\makeatother
\usepackage [ short ]{ optidef }
\makeatletter
\newcommand\semiHuge{\@setfontsize\semiHuge{23}{28}}
\makeatother

\DeclareMathOperator*{\E}{\mathbb{E}}

\usepackage{mathtools}
\DeclarePairedDelimiter\ceil{\lceil}{\rceil}
\DeclareMathAlphabet{\mathcal}{OMS}{cmsy}{m}{n}
\newcommand{\floor}[1]{\left\lfloor #1 \right\rfloor}
\newtheorem{definition}{Definition}
\newtheorem{theorem}{Theorem}

\newtheorem{lemma}{Lemma}

\pgfplotsset{compat=1.16}
\newtheorem{remark}{Remark}

\makeatletter
\newcommand{\thickhline}{%
    \noalign {\ifnum 0=`}\fi \hrule height 1pt
    \futurelet \reserved@a \@xhline
}
\newcolumntype{"}{@{\hskip\tabcolsep\vrule width 1pt\hskip\tabcolsep}}
\makeatother

\IEEEoverridecommandlockouts
\begin{document}
\title{Binary Modelling and Capacity-Approaching Coding for the IM/DD Channel}

\author{
\IEEEauthorblockN{Sarah~Bahanshal,~\IEEEmembership{Student Member,~IEEE,}
        Ahmad Abdel-Qader,~\IEEEmembership{Student Member,~IEEE,}
        Anas~Chaaban,~\IEEEmembership{Senior Member,~IEEE}}\\
 \IEEEauthorblockA{}       
        \thanks{The authors are with the School of Engineering, University of British Columbia, Kelowna, BC V1V1V7, Canada. 
        
        This paper is an extension of work presented in ISIT 2021 \cite{ourpaper}, and was funded by the Natural Science and Engineering Research Council (NSERC) under grant RGPIN-2018-04254.}}
 
%\IEEEpubid{\begin{minipage}[t]{\textwidth}\ \\[8pt]
%        {xxx-x-xxxx-xxxx-x \copyright 2020 IEEE}
%\end{minipage}}
\maketitle

\vspace{-1cm}

\begin{abstract}
The paper provides a new perspective on peak- and average-constrained Gaussian channels. Such channels model optical wireless communication (OWC) systems which employ intensity-modulation with direct detection (IM/DD). First, the paper proposes a new, capacity-preserving vector binary channel (VBC) model, consisting of dependent binary noisy bit-pipes. Then, to simplify coding over this VBC, the paper proposes coding schemes with varying levels of complexity, building on the capacity of binary-symmetric channels (BSC) and channels with state. The achievable rates are compared to capacity and capacity bounds, showing that coding for the BSC with state over the VBC achieves rates close to capacity at moderate to high signal-to-noise ratio (SNR), whereas simpler schemes achieve lower rates at lower complexity. The presented coding schemes are realizable using capacity-achieving codes for binary-input channels, such as polar codes. Numerical results are provided to validate the theoretical results and demonstrate the applicability of the proposed schemes.
\end{abstract}

\section{Introduction}
\label{secintro}
The increased number of users and the demand for data transfer increases the load on wireless networks. This leads to increased stress on the scarce radio frequency spectrum and motivates the search for other alternatives in the electromagnetic spectrum. One alternative that gained a renewed research interest in the last decade is the optical frequency spectrum (infrared, visible, and ultraviolet spectra), which can be used for wireless data transmission in what is known as optical wireless communication (OWC). OWC offers several advantages; arguably the most important is its unlicensed and vast spectrum, in particular in the infrared and visible light spectra \cite{hew_opticalwirelesscomp}. Additionally, OWC provides economic gains (low-cost deployment and infrastructure e.g.) and fast data rates. Typical applications include free-space optical communication for front-haul and back-haul links, and visible-light communications for indoor connectivity  \cite{chaaban2020tutorial,ku_fsosurvey}.

In OWC, information transmission is done by using light, and can be realized in two ways. One way is coherent communication, which is realized by modulating the phase and amplitude of the optical carrier similar to radio frequency communication. The other is incoherent communication which is simpler in practice, realized by modulating the intensity of a light source acting as a transmitter (LED or LASER), and detecting the incident light intensity using a photodetector as a receiver. This scheme is commonly known as intensity-modulation with direct detection (IM/DD). In IM/DD, the emitted light intensity is a nonnegative quantity with a peak constraint in addition to a possible average constraint for eye safety and practical operation reasons \cite{hwnc_opticalstory,ku_fsosurvey,ch_ptwiresur,
hew_opticalwirelesscomp,AdvancedOpticalWireless}. The received light intensity is corrupted by several noise sources, e.g., thermal noise and ambient light, often modelled as independent Gaussian noise \cite{ku_fsosurvey,chaaban2020tutorial,
book_capacityofopticalchannels}. This model is accurate under high intensity shot noise from ambient light \cite{WIinfraredCom,lmw_bounds,fh_signalling}.

The resulting additive Gaussian noise channel with a nonnegative, peak, and average constrained input is known as the IM/DD channel \cite{FaridHranilovic,chaaban2020tutorial}. A closed-form capacity expression for the IM/DD channel is not known to-date. However, several works derived capacity upper and lower bounds and asymptotic capacity results  \cite{lmw_bounds,FaridHranilovic,ChaabanMorvanAlouini,
bounds_mckellips,bounds_Thangaraj,bounds_clerckx}. Moreover, it is known that the capacity achieving input distribution is discrete with a finite number of mass points \cite{smith_finitemasspoints,chk_discreteinputdist}. Nonetheless, finding the optimal distribution remains a hard problem as it requires a maximization over the space of discrete input distributions. To simplify the search, \cite{fh_signalling} proposes a near-optimal input distribution by optimizing a discrete distribution with equal mass-point spacing. However, even if the optimal (or near-optimal) input distribution is found, random-coding using this distribution may not be appealing in practice. Instead, it is more practical to use binary codes like low-density parity-check codes (LDPC) \cite{ldpc_gallager} or polar codes \cite{arikan_channelpolarization,
erdal_ontheoriginofpolarcodes}. Motivated by this, it is interesting to investigate how closely the capacity can be approached using binary codes, especially codes for binary input and binary output (BIBO) channels.

Employing binary codes in a continuous channel decomposes it into a set of bit-pipes. This was investigated for achieving/approaching capacity using multi-level coding (MLC) and multi-stage decoding (MSD) \cite{wfh_mlc}, approximating the capacity of wireless networks using the linear-deterministic model (LDM) \cite{adt_deterministic}, and improving the reliability of transmission over fading channels \cite{BICM}. The same concept was applied in the IM/DD channel \cite{fh_signalling} to propose a capacity-approaching LDPC-based MLC/MSD scheme combined with a custom-designed mapper which produces a desired input distribution from the binary codeword symbols \cite{multilevel_nonuni,book_gallager1968information}. MLC/MSD over the IM/DD channel was also studied in \cite{tw_C}, which demonstrates the benefits of using probabilistic shaping through a proposed multi-level polar coded modulation scheme. MSD in this context entails decoding from the continuous channel output \cite{fh_signalling,multilevel_nonuni,
newmultilevel_ih,wfh_mlc,c_power_bwefficient_ldpc,
d_performance_ldpc,tw_C}, which prevents using BIBO channel codes (BCC). Decoding from a binarized output is possible, but leads to performance loss \cite[Sec. VII]{wfh_mlc}. Note that this latter option was not previously studied in the IM/DD channel context. 

From another perspective, the LDM \cite{adt_deterministic} provides an approximate binary decomposition of the additive white Gaussian noise channel (AWGN). It applies several approximations\footnote{such as replacing the power constraint and Gaussian noise by a peak constraint and peak constrained noise, and ignoring carry-overs in binary addition.} to decompose the AWGN channel into a set of independent noiseless bit-pipes, in addition to noisy bit-pipes that are discarded. The simplicity of the LDM enabled a better understanding of wireless networks and led to several advancements in network information theory \cite{st_feedbackcap,ht_interferencemitigation,CIT-042}. It is interesting to investigate whether the approximations in \cite{adt_deterministic} can be avoided to obtain a more accurate binary decomposition of the channel. This is indeed possible and it leads to an interesting connection with MLC/MSD as shown in this paper. 

This work is inspired by \cite{adt_deterministic,wfh_mlc}. It aims to develop an accurate binary decomposition of the IM/DD channel, and to propose coding schemes that only require BCC. To this end, the paper develops the vector binary channel (VBC) decomposition, which models the IM/DD channel as $N$ {\it noisy bit-pipes}, while carefully accounting for interactions between the transmit signal, noise, and carry-over bits. Each bit-pipe is a binary symmetric channel (BSC) with a cross-over probability that depends on the Gaussian noise distribution. By factoring the interactions between bit-pipes into the VBC model, capacity is preserved.

Then, the paper proposes coding schemes for the obtained VBC, all of which rely on independent encoding with different decoding schemes. The focus will be on decoding from the lowest bit-pipe (bit-pipe $0$) and proceeding successively to the highest one (bit-pipe $N-1$), which allows subtracting estimated carry-over bits from higher bit-pipes. First, a simple \textit{independent-decoding} (ID) scheme is described, which can be easily implemented using BCC, but exhibits a gap to capacity. The gap arises because the scheme ignores dependence between bit-pipes. To take this dependence into account, a superior {\it state-assisted decoding} (SD) scheme is proposes, which treats the bit-pipes as BSCs with state \cite[Ch.~7]{ElgamalKim}, where the state is the binary noise estimated from decoded bit-pipes. This scheme decreases the gap to capacity at moderate to high SNR. This scheme is further simplified by using the channel state to judiciously flip the output of a bit-pipe before discarding the state, leading to a BSC where BCC can be used. The achievable rate of this SD-BSC scheme is close to that of the SD scheme.

These schemes exhibit a gap to capacity at low SNR, where some information gets `pushed' to higher bit-pipes as  carry-over bits. To recover this information, a {\it carry-over-assisted decoding} (CD) scheme is proposed, which decodes bits sent over bit-pipe $i$ using output bits from bit-pipes $i,i+1,\ldots,i+q$ jointly (for some $q$). This scheme outperforms the ID and SD schemes at low SNR. A simplification of this scheme which can be implemented using BCC is also proposed.

Although the proposed VBC model and coding schemes are new, and despite yielding good performance, the proposed schemes have some limitations in terms of encoding and decoding. The proposed encoding constrains the input distribution to have $2^N$ equally spaced mass points with a distribution which is a convolution of Bernoulli distributions, which sacrifices some performance. Potential improvements include using a custom designed mapper or distribution matching as in \cite{fh_signalling,tw_A}, or using probabilistic shaping which has demonstrated its value in IM/DD channels \cite{prob_4pam,tw_A,tw_C}. For instance, \cite{prob_4pam} proposes a probabilistically shaped LDPC-coded $4$-level pulse-amplitude modulation ($4$-PAM) signalling for a peak-constrained IM/DD channel and tests it experimentally, \cite{tw_A} develops a probabilistic amplitude shaping scheme for $M$-PAM and shows its benefits for $6$-PAM in a certain range of SNR, and \cite{tw_B} validates \cite{tw_A} experimentally. On the other hand, the proposed decoding allows the decoder to benefit from dependencies between bit-pipe $i$ and lower bit-pipes, or between bit-pipe $i$ and higher bit-pipes, but not both. An improvement may be possible by combining the proposed SD and CD schemes. These issues are discussed towards the end of the paper.

Finally, the paper validates the theoretical results by implementing the ID and the SD-BSC schemes using polar codes, and evaluates their performance in terms of error rate and achievable rate. The simulations demonstrate that the theoretical results can be approached in practice. 

The results of the paper apply under a peak intensity constraint and also under both peak and average intensity constraints. They can be easily extended to the peak-constrained AWGN channel with a power (second moment) constraint, which is common in radio-frequency communications. 

The rest of the paper is organized as follows. Sec. \ref{sec_channelmodelimdd} presents the IM/DD channel model. Sec. \ref{sec_vbc} presents the decomposition of the IM/DD channel into the VBC, and shows that this decomposition is capacity-preserving. Sec. \ref{sec_codingschemes} discusses coding over the obtained VBC and connections with MLC/MSD. Sec. \ref{Sec:BCC} presents the proposed coding schemes and derives their achievable rates. Sec. \ref{sec_resultsanddiscussion} discusses the results and provides simulations using polar codes to validate the theoretical results. Sec. \ref{sec_conclusion} concludes the paper.

\section{The Peak-Constrained IM/DD Channel}
\label{sec_channelmodelimdd}
Consider an OWC system where optical intensity is used to send information from a transmitter to a receiver. The transmitted signal is represented by a random variable $X$, which must satisfy nonnegativity, peak, and average intensity constraints (for safety and practical reasons). This variable $X$ represents the transmitted optical intensity.\footnote{or the electric current used to generate the optical signal, assuming operation in the linear regime of the solid-state light emitting device where the relation between intensity $X$ and current $I$ can be approximated as $X=\eta I$ with $\eta$ being the electrical-to-optical conversion efficiency of the light source \cite{ILX_LD,ch_ptwiresur}.} The received signal is represented by a random variable $Y$ which combines electric current contributions from the received signal intensity, shot noise, ambient light, and thermal noise. Several channel models can be considered to model this OWC as described below.

The most basic channel model for OWC is the Poisson channel \cite{d_poissonchannel}, where $Y$ is nonnegative. However, if the number of received photons is large, the Poisson channel can be approximated as a Gaussian channel with input dependent noise \cite{Moser,Safari_DependentNoise,chaaban2020tutorial}. Additionally, if thermal noise and noise from ambient light dominate noise due to transmitted photon fluctuations, this Gaussian noise can be approximated as input-independent \cite{WIinfraredCom,lmw_bounds,fh_signalling}. Combining the Gaussian noise to the received signal results in $Y$ which can be negative. In this paper, the latter channel model is considered, which is denoted an IM/DD channel for brevity henceforth, defined as follows.

\begin{definition} [IM/DD Channel]\label{def_imddchannel} The IM/DD channel is defined by the input-output relation
\begin{align}
\label{eq_imdd}
Y= X + Z,
\end{align}
where $Z$ is Gaussian with zero mean and variance $\sigma^2=1$ (without loss of generality) which is independent and identically distributed (i.i.d.) over time, and where $X\geq 0$ is constrained by a peak constraint $X \leq A$ and an average constraint $\E[X]\leq E$.
\end{definition}

The ratio between the average and the peak intensity constraints is denoted by $\rho =\frac{E}{A}$, where $0<\rho\leq1$. The capacity bounds of the IM/DD channel depend on $\rho$. If $\rho \geq \frac{1}{2}$, then the average intensity constraint is said to be inactive, where setting $\mathbb{E}[X]=A/2$ achieves capacity \cite{lmw_bounds,ChaabanMorvanAlouini}. However, if $\rho<1/2$, then capacity depends on $E$ and is achieved by setting $E[X]=E$.

The transmitter wants to send a message $m \in \{1,\ldots,2^{nR}\}$ (assuming $2^{nR}$ is an integer) with rate $R$ over $n$ channel uses. Encoding, decoding, achievable rate, and capacity are defined in the standard Shannon sense \cite{cover_elementsIT2}. The capacity is the highest achievable rate given by
\begin{align}
\label{eq_capimdd}
C^{\text{IM/DD}}= \max_{P_X} I(X;Y) 
\end{align}
where $P_X$ is the input distribution and $I(X;Y)$ is the mutual information between $X$ and $Y$. There is no closed-form expression for $C^{\rm IM/DD}$ to-date, yet it is computable numerically using optimization over the input distribution \cite{chk_discreteinputdist}. 
 %As discussed in Sec. \ref{secintro}, t
 This paper aims to design and analyze a coding scheme that approaches this capacity via decomposing the channel into simpler binary channels. To this end, a binary decomposition of the IM/DD channel is proposed in the following section.

\section{A Vector Binary Channel Model}
\label{sec_vbc}
To aid in designing practical coding schemes for the IM/DD channel using BCC, we start by decomposing the IM/DD channel into bit-pipes in a manner similar to \cite{adt_deterministic} but with the following differences. We model the interaction between bit-pipes through carry-over bits, and we transform the Gaussian noise into binary noise which affects all bit-pipes. This leads to noisy bit-pipes, contrary to \cite{adt_deterministic} which leads to a noiseless approximation.

\subsection{Mathematical Formulation}
\label{sec_mathform}
To decompose the IM/DD channel in Def. \ref{def_imddchannel} into a set of bit-pipes, let us represent $X$ and $Y$ using a nonnegative $N$-bit representation. This can only represent numbers in $\{0,1,\ldots,2^N-1\}$. Since $Y$ is unbounded (as the noise $Z$ is Gaussian), it should be truncated before converting it to binary. To reduce the distortion caused by this truncation, let us introduce a margin of width $\beta >0$ around $[0,A]$ and treat a received $Y$ outside this interval as an erasure. Formally, let $\tilde{Y}$
\begin{align}
\label{eq_shifted}
\tilde{Y}= X + Z + \beta,
\end{align}
for some $\beta>0$. Then, the truncated signal is defined as
\begin{align}
\label{eq_shifted_truncated}
Y' = \begin{cases}
        \tilde{Y}, & \text{for } \tilde{Y} \in [0, A+2\beta], \\
        \xi, & \text{otherwise},
        \end{cases}
\end{align}
where $\xi$ denotes an erasure.\footnote{We can clip the signal $\tilde{Y}$ at $0$ and $A+2\beta$ instead, leading to clipping errors. Since erasures are less expensive in terms of data rate than errors, we opt for erasing instead of clipping.} The erasure probability is thus given by $\epsilon = \mathcal{P}\{\tilde{Y}<0\}+\mathcal{P}\{\tilde{Y}>A+2\beta\}$ where $\mathcal{P}\{\cdot\}$ denotes the probability of an event. Note that $\epsilon$ depends on the input distribution $P_X$ which is unknown at this point, and is upper bounded by the worst-case scenario given by $\mathcal{P}\{\tilde{Y}<0|X=0\} + \mathcal{P}\{\tilde{Y}>A+2\beta|X=A\}$, leading to
\begin{align}
\label{eq_perasure}
\epsilon \leq \bar{\epsilon}\triangleq 2Q(\beta),
\end{align}
where $Q(x) = \frac{1}{\sqrt{2\pi}} \int_x^\infty e^{-\frac{u^2}{2}}{\rm d}u$ is the standard Gaussian tail distribution function. Note that $\bar{\epsilon} \rightarrow 0$ as $\beta$ grows.\footnote{Choosing $\beta\geq 3$ suffices to ensure negligible $\epsilon$, since the added noise is standard Gaussian, and $\mathcal{P}\{Z\in[-\beta,\beta]\}\approx 0.99$.} Now, the processed channel output can have one of two states, either $Y' = \tilde{Y}$ or $Y^\prime$ is an erasure $\xi$. Next, we focus on unerased outputs, i.e., $Y'\neq\xi$.

\subsubsection{Modelling Unerased Outputs ($\tilde{Y}\in[0,A+2\beta]$)}
\label{sec_unerased}
Throughout this subsection (in \eqref{eq_y hat}-\eqref{Zib}), we focus on $Y'=\tilde{Y}$, with the understanding that erasures will be reintegrated into the model later. Starting with $Y' =\tilde{Y}$ in \eqref{eq_shifted_truncated}, let
\begin{align}
\label{eq_y hat}
\hat{Y} = \gamma {Y'},
\end{align}
for some $\gamma>0$, which determines the resolution of the binary representation. Then, let the number of bits used to quantize the channel output be 
\begin{align}
\label{Eq:N}
N= \ceil{\log_2{\gamma(A+2\beta)}},
\end{align} 
since $\hat{Y}$ is in $[0,\gamma(A+2\beta)]$. Note that a large $\gamma$ leads to a more accurate binary representation of $Y'$. The received signal is then represented as a weighted sum of $N$ bits as $\hat{Y} \approx \sum_{i=0}^{N-1} Y_i^b 2^i$, with the difference being the quantization distortion, where 
\begin{align}
\label{eq_y bin}
Y_i^b  =\floor{ \frac{\hat{Y}}{2^i} } \bmod 2,\; \; i=0,1,\ldots,N-1,
\end{align}
is the binary representation of $\hat{Y}$ ($\lfloor\cdot\rfloor$ is the floor function). Substituting \eqref{eq_shifted} and \eqref{eq_y hat} in \eqref{eq_y bin} and using the relation $\floor{a+b}=\floor{a}+\floor{b}+\floor{a-\floor{a}+b-\floor{b}}$ leads to
\begin{align}
\label{eq_y_bin_1}
Y_i^b & = \floor{ \frac{\gamma(X+Z+\beta)}{2^i} } \bmod{2}=  \left(\floor{ \frac{\gamma {X}}{2^{i}} }+\floor{ \frac{\gamma(Z +\beta)}{2^{i}} }+ \floor{\delta_i} \right) \bmod{2},
\end{align}
where $\delta_i$, $i=0,1\ldots,N-1$, is the sum of the fractional parts of $\frac{\gamma X}{2^i}$ and $\frac{\gamma(Z+\beta)}{2^i}$, given by
\begin{align}
\label{eq_delta}
\delta_i =  \frac{\gamma X}{2^i} - \floor{ \frac{\gamma X}{2^i} } + \frac{\gamma(Z+\beta)}{2^i} - \floor{ \frac{\gamma(Z+\beta)}{2^i} }.
\end{align}

The next goal is to relate $\gamma X$ and $\gamma(Z+\beta)$ \eqref{eq_y_bin_1} to their binary representation, to obtain a fully binary model (binary input, output, and noise). Two cases are considered next: $\gamma(Z+\beta)\geq0$ and $\gamma(Z+\beta)<0$, because the binary representation when $\gamma(Z+\beta)<0$ requires special treatment.

\paragraph{Nonnegative $\gamma(Z+\beta)$} In this case, 
\begin{align}
\label{eq_y bin 3}
Y_i^b  = \left( \floor{\frac{\hat{X}}{2^{i}}} \bmod{2}+\floor{ \frac{\hat{Z}}{2^{i}} } \bmod{2}+\floor{\delta_i} \bmod{2} \right) \bmod{2},
\end{align}
which follows since $(a+b)\bmod 2= (a \bmod2 +b \bmod 2) \bmod2$, where 
\begin{align}
\label{eq_xzhat}
    \hat{X}  &=\gamma X\\
    \hat{Z}  &= \gamma \left( Z+ \beta \right) \text{ (given $\gamma(Z+\beta)\geq0$)}.
\end{align}
Note that $\hat{X} \approx \sum_{i=0}^{N-1} X_i^b 2^i$ and $\hat{Z} \approx \sum_{i=0}^{N-1}  Z_i^b2^i$ for $i=0,1,\ldots,N-1$, where
\begin{align}
\label{eq_xbin}
X_i^b  &= \floor{ \frac{\hat{X}}{2^i} } \bmod{2},\ \
%\label{eq_zbin}
Z_i^b  = \floor{ \frac{\hat{Z}}{2^i} } \bmod{2},\ \text{and } \ 
%\label{eq_cbin}
W_i^b = \floor{ \delta_i } \bmod{2},
\end{align}
%, i.e, in terms of \eqref{eq_xbin} and \eqref{eq_zbin} as follows
%It remains to binarize $\delta_i$, let $W_i^b$ be given by
Consequently, using \eqref{eq_xbin} in \eqref{eq_y bin 3} yields
\begin{align}
\label{eq_yxzc}
Y_i^b  = (X_i^b  + Z_i^b + W_i^b )\bmod{2}, \ i=0,\ldots,N-1.
\end{align}
This representation connects the binary representation of $\hat{Y}$ (i.e., $Y_i^b$) with the binary representation of $\hat{X}$ and $\hat{Z}$ (i.e., $X_i^b$ and $Z_i^b$), and a bit $W_i^b$ which one can show to be equal to
\begin{align}
\label{eq_cbin2}
W_i^b = (X_{i-1}^b Z_{i-1}^b + X_{i-1}^b W_{i-1}^b+ Z_{i-1}^b W_{i-1}^b) \bmod{2},
\end{align}
for $i=1,\ldots,N-1$, i.e., $W_i^b$ is the carry-over bit resulting from the modulo-$2$ addition of `lower significance' bits $X_{i-1}^b$, $Z_{i-1}^b$, and $W_{i-1}^b$. For $i=0$, the carry-over is defined as $W_0^b=0$. This can be obtained from \eqref{eq_cbin2} by defining $X_{i}^b=0$ for all $i<0$ (via choosing $\hat{X}$ to be an integer) which will be used in Sec. \ref{Sec:BCC}. Next, we discuss the other case where $\gamma(Z+\beta)<0$.

\paragraph{Negative $\gamma(Z+\beta)$}
In this case, using $\hat{Z}=\gamma(Z+\beta)$ in \eqref{eq_xbin} results in a discrepancy between $Y_i^b$ in \eqref{eq_y_bin_1} and \eqref{eq_yxzc} as a result of the relation $(-a)\bmod2= (a)\bmod2 $. For example, let $X=A=4$, $\gamma=1$, $\beta=2$, and $Z=-3$. Then, $\hat{X}=4$ and $\hat{Z}=-1$ using \eqref{eq_xzhat}, and $N=3$ using \eqref{Eq:N}. The binary representations of $\hat{X}$ and $\hat{Z}$ (using \eqref{eq_xbin}) from the least significant to most significant bit are $(X_0^b, X_1^b,X_2^b)=(0,0,1)$ and $(Z_0^b,Z_1^b,Z_2^b)=(1,0,0)$, respectively. The carry-over bits are $(W_0^b,W_1^b,W_2^b)=(0,0,0)$ using \eqref{eq_cbin2}. Then using \eqref{eq_yxzc} results in $(Y_0^b,Y_1^b,Y_2^b)=(1,0,1)$, i.e., $\hat{Y}=5$, which is incorrect since $\hat{Y}=\tilde{Y}=\gamma(X+Z+\beta)=3$. 

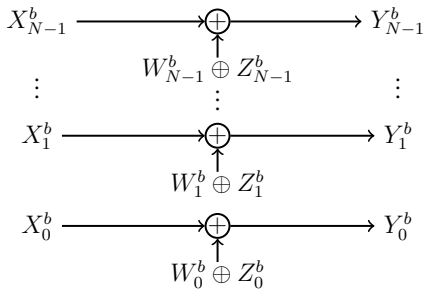
\begin{figure}
   \centering
\tikzset{every picture/.style={line width=0.75pt}} %set default line width to 0.75pt        
\begin{tikzpicture}[scale=0.8, every node/.style={scale=0.8}]
\node (x0) at (0,0) {$X^{b}_{0}$};
\node (n0) at (3,-.8) {$W^{b}_{0} \oplus Z^{b}_{0}$};
\node (y0) at (6,0) {$Y^{b}_{0}$};
\node[circle,draw] (p0) at (3,0) {};
\node at (p0) {$+$};
\draw[->] (x0) to (p0);
\draw[->] (p0) to (y0);
\draw[->] ($(n0.center)+(0,0.2)$) to (p0);

\node (x1) at (0,1.5) {$X^{b}_{1}$};
\node (n1) at (3,.7) {$W^{b}_{1} \oplus Z^{b}_{1}$};
\node (y1) at (6,1.5) {$Y^{b}_{1}$};
\node[circle,draw] (p1) at (3,1.5) {};
\node at (p1) {$+$};
\draw[->] (x1) to (p1);
\draw[->] (p1) to (y1);
\draw[->] ($(n1.center)+(0,0.2)$) to (p1);

\node (dx1) at (0,2.4) {$\vdots$};
\node (dn1) at (3,2.2) {$\vdots$};
\node (dy1) at (6,2.4) {$\vdots$};

\node (xN-1) at (0,3.4) {$X^{b}_{N-1}$};
\node (nN-1) at (3,2.6) {$W^{b}_{N-1} \oplus Z^{b}_{N-1}$};
\node (yN-1) at (6,3.4) {$Y^{b}_{N-1}$};
\node[circle,draw] (pN-1) at (3,3.4) {};
\node at (pN-1) {$+$};
\draw[->] (xN-1) to (pN-1);
\draw[->] (pN-1) to (yN-1);
\draw[->] ($(nN-1.center)+(0,0.2)$) to (pN-1);

\end{tikzpicture}
   \caption{The vector binary channel with $N$ bit-pipes, inputs $X_i^b$, $i=0,\ldots,N-1$, and noise being the modulo-$2$ sum of $Z_i^b$ (Bernoulli distributed) and $W_i^b$ (carry-over from bit-pipe~$i-1$).}
    \label{fig_model}
\end{figure}

This can be remedied by using the two's complement when $\gamma (Z+\beta)<0$ to define $\hat{Z}= 2^N+\gamma (Z+\beta)$. In the example above where $\gamma(Z+\beta)=-1<0$, this leads to $\hat{Z}=7$ whose binary representation is $(Z_0^b,Z_1^b,Z_2^b)=(1,1,1)$. Thus, $(Y_0^b,Y_1^b,Y_2^b)=(1,1,0)$,\footnote{Note that in this case there will be a carry-over bit $W_3^b=1$, but this is ignored since we use a $3$-bit representation ($N=3$).} i.e., $\hat{Y}=3$, as desired. Combining both cases of positive and negative $\gamma(Z+\beta)$ leads to the following generalization
\begin{align}
\label{eq_zhat}
\hat{Z} &= 2^N {\sf I}_{\gamma(Z+\beta)<0}+\gamma (Z+\beta),\\
\label{Zib}
Z_i^b&=\left\lfloor\frac{\hat{Z}}{2^i}\right\rfloor\bmod 2,\quad i=0,1,\ldots,N-1,
\end{align}
where ${\sf I}_{\mathcal{E}}$ is an indicator function for event $\mathcal{E}$, which equals $0$ when $\mathcal{E}$ is false and equals $1$ otherwise. This generalization will be used henceforth. Fig. \ref{fig_calculating noise dist} shows an example of this binarization of $\hat{Z}$, which is discussed in detail in Sec. \ref{sec_noise}. We denote the resulting representation as the VBC defined next (Fig. \ref{fig_model}).

\begin{definition}[Vector Binary Channel]
\label{Def:VBC}
The VBC is characterized by a collection of $N$ bit-pipes with an input-output relation $Y_i^b=(X_i^b  +Z_i^b +W_i^b )\bmod{2}$, where $Y_i^b$ is the $i$th bit-pipe's output bit, $X_i^b$ is the input bit, $Z_i^b$ is binary noise obtained from the $N$-bit binary representation of the biased and scaled Gaussian noise $\gamma(Z+\beta)$ \eqref{Zib}, and $W_i^b$ is a carry-over bit \eqref{eq_cbin2}. 
\end{definition}

Next, erasures are reintegrated into the definition of the resulting binary channels in \eqref{eq_yxzc}.

\subsubsection{Modelling Output with Erasures ($\tilde{Y}\notin[0,A+2\beta]$)}
Recall that erasures were ignored in the previous analysis. To model the IM/DD channel more accurately, the channel in \eqref{eq_yxzc} should be concatenated with a binary erasure channel (BEC) with erasure probability $\bar{\epsilon}$ \eqref{eq_perasure}. Let 
\begin{align}
\label{eq_imdd_erasuress}
    Y_i^c = \begin{cases}
        Y_i^b= (X_i^b  + Z_i^b + W_i^b )\bmod{2}, & \text{with probability\;\; } \;\;1-\bar{\epsilon}, \\
        \xi, & \text{with probability \;\;\;\;}\bar{\epsilon},
        \end{cases}
\end{align}
denote the output of the concatenated channel, for $i=0,\ldots, N-1$, where $X_i^b$, $W_i^b$, and $Z_i^b$ are defined in \eqref{eq_xbin}, \eqref{eq_cbin2}, and \eqref{Zib}, respectively. This leads to the VBC with erasures defined next.

\begin{definition} [VBC with erasures: $\text{VBC}_{\bar{\epsilon}}$]
\label{def_vbce}
$\text{VBC}_{\bar{\epsilon}}$ is a channel with inputs $X_i^b$,  $i=0,\ldots, N-1$, and outputs $Y_i^c$ as defined in \eqref{eq_imdd_erasuress}.
\end{definition}

The VBC with erasures defined above is used to approximate the IM/DD channel in \eqref{eq_imdd}, as a channel with a vector binary input, output (which could be erased), and noise. Let $\boldsymbol{X}^b=(X_0^b,\ldots,X_{N-1}^b )^T$ and $\boldsymbol{Y}^c=(Y_0^c,\ldots,Y_{N-1}^c )^T$ represent the binary input and output vectors, respectively. Also, let $P_{\boldsymbol{X}_i^b}$ be the distribution of $\boldsymbol{X}_i^b$. Then the following statement holds.

\begin{lemma}
\label{lem_capvbce}
The capacity of $\textnormal{VBC}_{\bar{\epsilon}}$ in Def. \ref{def_vbce} satisfies $C^{\textnormal{VBC}_{\bar{\epsilon}}} = \max_{P_{\boldsymbol{X}_i^b}} I(\boldsymbol{X}^b;\boldsymbol{Y}^c) < C^{\text{IM/DD}}$, where $P_{\boldsymbol{X}_i^b}$ satisfies the peak constraint $\gamma A$ and the average constraint $\gamma E$.
\end{lemma}
\begin{IEEEproof}
The equality follows from Shannon's channel coding theorem, while the inequality follows using the data processing inequality and noting that $\bar{\epsilon}>0$.
\end{IEEEproof}

It remains to express the distribution of $Z_i^b$. Note that when the output is not erased, the shifted and scaled Gaussian noise $\gamma(Z+\beta)$ becomes dependent on the input $X$ after truncating the output to be in $[0,A+2\beta]$. For instance, if $X=A$ and $\tilde{Y}\in[0,A+2\beta]$, then $Z$ must be in $[-(A+\beta),\beta]$, but if $X=0$ and $\tilde{Y}\in[0,A+2\beta]$, then $Z$ must be in $[-\beta,A+\beta]$. In the following section, this dependence is ignored, leading to an approximation that is accurate if $\beta\geq3$. Then, this approximation is used to express the distribution of noise $Z_i^b$.

\vspace{-0.2cm}

\subsection{Characterizing the Binary Noise Distribution}
\label{sec_noise}
To find $\mathcal{P}\{Z_i^b=1\}$, the noise $Z$ (given $Y' \neq \xi$) is assumed to be independent of $X$ by considering the interval $[-(A+\beta), A+\beta]$ as a unified support for $Z$ regardless of $X$ (from the smallest to the largest values of $Z$ for any $X$). If $\beta\geq 3$, this approximation results in a negligible difference, as $Q(\beta)\leq Q(3)= 0.0013$. Specifically, if $X=0$, $ Z \geq -\beta$ is replaced by $Z \geq-(A+\beta)$ by the assumption above. This does not significantly change the distribution of $Z$ if $\beta\geq 3$, since the `added noise density' in $[-(A+\beta),-\beta]$ is less than $Q(\beta)\leq Q(3)$.   
\begin{figure}[t]
    \centering
    \tikzset{every picture/.style={scale=.8}, every node/.style={scale=.9}}
%         \input{noise_dist}

% Minimal pgfplots version: 1.3
%  http://www.mathworks.com/matlabcentral/fileexchange/22022-matlab2tikz
\begin{tikzpicture}

\begin{axis}[%
width=9in,
height=1.5in,
xmin=-2,
xmax=4,
xlabel={$\gamma(Z+\beta)$},
xmajorgrids,
ymin=0,
ymax=0.4,
ylabel={$P_{\gamma(Z+\beta)}$},
ymajorgrids,
legend style={legend cell align=left,align=left,draw=white!15!black}
]
\addplot [color=black,solid,forget plot,line width=1pt]
  table[row sep=crcr]{%
-2	0.00443184841193801\\
-1.99	0.00456658995467014\\
-1.98	0.00470495752693398\\
-1.97	0.00484703290597895\\
-1.96	0.00499289921361238\\
-1.95	0.00514264092305394\\
-1.94	0.00529634386531102\\
-1.93	0.00545409523505656\\
-1.92	0.00561598359599097\\
-1.91	0.00578209888566947\\
-1.9	0.00595253241977585\\
-1.89	0.00612737689582369\\
-1.88	0.00630672639626593\\
-1.87	0.00649067639099336\\
-1.86	0.00667932373920262\\
-1.85	0.00687276669061397\\
-1.84	0.00707110488601945\\
-1.83	0.00727443935714122\\
-1.82	0.00748287252578055\\
-1.81	0.00769650820223732\\
-1.8	0.00791545158297997\\
-1.79	0.00813980924754602\\
-1.78	0.00836968915465302\\
-1.77	0.00860520063749967\\
-1.76	0.00884645439823723\\
-1.75	0.00909356250159105\\
-1.74	0.00934663836761228\\
-1.73	0.00960579676353959\\
-1.72	0.00987115379475114\\
-1.71	0.0101428268947871\\
-1.7	0.0104209348144226\\
-1.69	0.0107055976097722\\
-1.68	0.0109969366294056\\
-1.67	0.0112950745004561\\
-1.66	0.0116001351137026\\
-1.65	0.0119122436076052\\
-1.64	0.012231526351278\\
-1.63	0.0125581109263782\\
-1.62	0.0128921261078953\\
-1.61	0.0132337018438214\\
-1.6	0.0135829692336856\\
-1.59	0.0139400605059358\\
-1.58	0.0143051089941497\\
-1.57	0.01467824911206\\
-1.56	0.0150596163273774\\
-1.55	0.0154493471343952\\
-1.54	0.0158475790253608\\
-1.53	0.0162544504606005\\
-1.52	0.0166701008373811\\
-1.51	0.017094670457497\\
-1.5	0.0175283004935685\\
-1.49	0.0179711329540396\\
-1.48	0.018423310646862\\
-1.47	0.0188849771418562\\
-1.46	0.019356276731737\\
-1.45	0.0198373543917953\\
-1.44	0.0203283557382258\\
-1.43	0.0208294269850922\\
-1.42	0.0213407148999228\\
-1.41	0.0218623667579294\\
-1.4	0.0223945302948429\\
-1.39	0.0229373536583607\\
-1.38	0.0234909853582014\\
-1.37	0.024055574214763\\
-1.36	0.0246312693063825\\
-1.35	0.0252182199151944\\
-1.34	0.0258165754715877\\
-1.33	0.0264264854972617\\
-1.32	0.0270480995468818\\
-1.31	0.0276815671483366\\
-1.3	0.0283270377416012\\
-1.29	0.0289846606162094\\
-1.28	0.0296545848473413\\
-1.27	0.0303369592305316\\
-1.26	0.0310319322150083\\
-1.25	0.0317396518356674\\
-1.24	0.0324602656436974\\
-1.23	0.0331939206358611\\
-1.22	0.0339407631824492\\
-1.21	0.0347009389539188\\
-1.2	0.0354745928462314\\
-1.19	0.0362618689049062\\
-1.18	0.0370629102478065\\
-1.17	0.0378778589866775\\
-1.16	0.0387068561474556\\
-1.15	0.0395500415893702\\
-1.14	0.0404075539228603\\
-1.13	0.0412795304263304\\
-1.12	0.0421661069617703\\
-1.11	0.0430674178892657\\
-1.1	0.0439835959804272\\
-1.09	0.0449147723307671\\
-1.08	0.0458610762710549\\
-1.07	0.0468226352776832\\
-1.06	0.047799574882077\\
-1.05	0.0487920185791828\\
-1.04	0.0498000877350708\\
-1.03	0.0508239014936912\\
-1.02	0.0518635766828206\\
-1.01	0.0529192277192403\\
-1	0.0539909665131881\\
-0.99	0.0550789023721258\\
-0.98	0.056183141903868\\
-0.97	0.0573037889191171\\
-0.96	0.0584409443334515\\
-0.95	0.0595947060688161\\
-0.94	0.0607651689545648\\
-0.93	0.0619524246281052\\
-0.92	0.0631565614351987\\
-0.91	0.0643776643299694\\
-0.9	0.0656158147746766\\
-0.89	0.0668710906393072\\
-0.88	0.0681435661010446\\
-0.87	0.0694333115436742\\
-0.86	0.0707403934569834\\
-0.85	0.072064874336218\\
-0.84	0.0734068125816569\\
-0.83	0.0747662623983676\\
-0.82	0.0761432736962073\\
-0.81	0.077537891990134\\
-0.8	0.0789501583008941\\
-0.79	0.0803801090561542\\
-0.78	0.0818277759921428\\
-0.77	0.0832931860558745\\
-0.76	0.0847763613080222\\
-0.75	0.0862773188265115\\
-0.74	0.0877960706109056\\
-0.73	0.089332623487655\\
-0.72	0.0908869790162829\\
-0.71	0.0924591333965807\\
-0.7	0.0940490773768869\\
-0.69	0.095656796163524\\
-0.68	0.0972822693314675\\
-0.67	0.0989254707363237\\
-0.66	0.100586368427691\\
-0.65	0.102264924563978\\
-0.64	0.103961095328764\\
-0.63	0.105674830848764\\
-0.62	0.107406075113484\\
-0.61	0.109154765896647\\
-0.6	0.110920834679456\\
-0.59	0.112704206575771\\
-0.58	0.114504800259292\\
-0.57	0.116322527892807\\
-0.56	0.118157295059582\\
-0.55	0.120009000696986\\
-0.54	0.121877537032402\\
-0.53	0.123762789521523\\
-0.52	0.125664636789088\\
-0.51	0.127582950572142\\
-0.5	0.129517595665892\\
-0.49	0.131468429872231\\
-0.48	0.133435303951002\\
-0.47	0.135418061574071\\
-0.46	0.137416539282282\\
-0.45	0.13943056644536\\
-0.44	0.141459965224839\\
-0.43	0.143504550540062\\
-0.42	0.145564130037348\\
-0.41	0.147638504062356\\
-0.4	0.149727465635745\\
-0.39	0.151830800432162\\
-0.38	0.153948286762634\\
-0.37	0.156079695560421\\
-0.36	0.158224790370383\\
-0.35	0.16038332734192\\
-0.34	0.162555055225534\\
-0.33	0.164739715373077\\
-0.32	0.166937041741714\\
-0.31	0.169146760901672\\
-0.3	0.171368592047807\\
-0.29	0.173602247015033\\
-0.28	0.175847430297662\\
-0.27	0.178103839072694\\
-0.26	0.18037116322708\\
-0.25	0.182649085389022\\
-0.24	0.184937280963305\\
-0.23	0.18723541817073\\
-0.22	0.18954315809164\\
-0.21	0.191860154713599\\
-0.2	0.194186054983213\\
-0.19	0.196520498862137\\
-0.18	0.198863119387276\\
-0.17	0.201213542735197\\
-0.16	0.203571388290759\\
-0.15	0.205936268719975\\
-0.14	0.208307790047108\\
-0.13	0.210685551736015\\
-0.12	0.213069146775718\\
-0.11	0.21545816177022\\
-0.0999999999999999	0.217852177032551\\
-0.0899999999999999	0.220250766683033\\
-0.0800000000000001	0.222653498751761\\
-0.0700000000000001	0.22505993528527\\
-0.0600000000000001	0.227469632457386\\
-0.05	0.229882140684233\\
-0.04	0.232297004743366\\
-0.03	0.234713763897012\\
-0.02	0.23713195201938\\
-0.01	0.239551097728013\\
0	0.241970724519143\\
0.0100000000000002	0.244390350907\\
0.02	0.246809490567043\\
0.0300000000000002	0.249227652483066\\
0.04	0.251644341098117\\
0.0499999999999998	0.254059056469189\\
0.0600000000000001	0.25647129442562\\
0.0699999999999998	0.258880546731149\\
0.0800000000000001	0.261286301249553\\
0.0899999999999999	0.263688042113818\\
0.1	0.266085249898755\\
0.11	0.268477401797002\\
0.12	0.270863971798338\\
0.13	0.273244430872216\\
0.14	0.275618247153457\\
0.15	0.277984886130996\\
0.16	0.280343810839621\\
0.17	0.28269448205458\\
0.18	0.285036358489007\\
0.19	0.287368896994028\\
0.2	0.289691552761483\\
0.21	0.292003779529141\\
0.22	0.294305029788325\\
0.23	0.296594754993816\\
0.24	0.298872405775953\\
0.25	0.301137432154804\\
0.26	0.3033892837563\\
0.27	0.30562741003021\\
0.28	0.307851260469853\\
0.29	0.310060284833416\\
0.3	0.312253933366761\\
0.31	0.314431657027597\\
0.32	0.316592907710893\\
0.33	0.318737138475402\\
0.34	0.320863803771172\\
0.35	0.322972359667914\\
0.36	0.325062264084082\\
0.37	0.327132977016554\\
0.38	0.329183960770765\\
0.39	0.331214680191153\\
0.4	0.3332246028918\\
0.41	0.335213199487106\\
0.42	0.337179943822381\\
0.43	0.339124313204192\\
0.44	0.341045788630353\\
0.45	0.342943855019384\\
0.46	0.344818001439333\\
0.47	0.346667721335792\\
0.48	0.348492512758974\\
0.49	0.350291878589726\\
0.5	0.3520653267643\\
0.51	0.35381237049778\\
0.52	0.355532528505997\\
0.53	0.357225325225801\\
0.54	0.358890291033545\\
0.55	0.360526962461648\\
0.56	0.362134882413092\\
0.57	0.363713600373713\\
0.58	0.365262672622154\\
0.59	0.366781662437336\\
0.6	0.368270140303323\\
0.61	0.369727684111432\\
0.62	0.371153879359466\\
0.63	0.372548319347933\\
0.64	0.373910605373128\\
0.65	0.375240346916938\\
0.66	0.376537161833254\\
0.67	0.377800676530865\\
0.68	0.379030526152702\\
0.69	0.380226354751325\\
0.7	0.381387815460524\\
0.71	0.382514570662924\\
0.72	0.383606292153479\\
0.73	0.384662661298743\\
0.74	0.385683369191816\\
0.75	0.386668116802849\\
0.76	0.387616615125014\\
0.77	0.388528585315836\\
0.78	0.38940375883379\\
0.79	0.390241877570074\\
0.8	0.391042693975456\\
0.81	0.391805971182121\\
0.82	0.392531483120429\\
0.83	0.393219014630497\\
0.84	0.393868361568541\\
0.85	0.394479330907889\\
0.86	0.395051740834611\\
0.87	0.395585420837687\\
0.88	0.396080211793656\\
0.89	0.396535966045686\\
0.9	0.396952547477012\\
0.91	0.397329831578688\\
0.92	0.397667705511609\\
0.93	0.397966068162751\\
0.94	0.398224830195607\\
0.95	0.398443914094764\\
0.96	0.398623254204605\\
0.97	0.3987627967621\\
0.98	0.398862499923666\\
0.99	0.398922333786082\\
1	0.398942280401433\\
1.01	0.398922333786082\\
1.02	0.398862499923666\\
1.03	0.3987627967621\\
1.04	0.398623254204605\\
1.05	0.398443914094764\\
1.06	0.398224830195607\\
1.07	0.397966068162751\\
1.08	0.397667705511609\\
1.09	0.397329831578688\\
1.1	0.396952547477012\\
1.11	0.396535966045686\\
1.12	0.396080211793656\\
1.13	0.395585420837687\\
1.14	0.395051740834611\\
1.15	0.394479330907889\\
1.16	0.393868361568541\\
1.17	0.393219014630497\\
1.18	0.392531483120429\\
1.19	0.391805971182121\\
1.2	0.391042693975456\\
1.21	0.390241877570074\\
1.22	0.38940375883379\\
1.23	0.388528585315836\\
1.24	0.387616615125014\\
1.25	0.386668116802849\\
1.26	0.385683369191816\\
1.27	0.384662661298743\\
1.28	0.383606292153479\\
1.29	0.382514570662924\\
1.3	0.381387815460524\\
1.31	0.380226354751325\\
1.32	0.379030526152702\\
1.33	0.377800676530865\\
1.34	0.376537161833254\\
1.35	0.375240346916938\\
1.36	0.373910605373128\\
1.37	0.372548319347933\\
1.38	0.371153879359466\\
1.39	0.369727684111432\\
1.4	0.368270140303323\\
1.41	0.366781662437336\\
1.42	0.365262672622154\\
1.43	0.363713600373713\\
1.44	0.362134882413092\\
1.45	0.360526962461648\\
1.46	0.358890291033545\\
1.47	0.357225325225801\\
1.48	0.355532528505997\\
1.49	0.35381237049778\\
1.5	0.3520653267643\\
1.51	0.350291878589726\\
1.52	0.348492512758974\\
1.53	0.346667721335792\\
1.54	0.344818001439333\\
1.55	0.342943855019384\\
1.56	0.341045788630353\\
1.57	0.339124313204192\\
1.58	0.337179943822381\\
1.59	0.335213199487106\\
1.6	0.3332246028918\\
1.61	0.331214680191153\\
1.62	0.329183960770765\\
1.63	0.327132977016554\\
1.64	0.325062264084082\\
1.65	0.322972359667914\\
1.66	0.320863803771172\\
1.67	0.318737138475402\\
1.68	0.316592907710893\\
1.69	0.314431657027597\\
1.7	0.312253933366761\\
1.71	0.310060284833416\\
1.72	0.307851260469853\\
1.73	0.30562741003021\\
1.74	0.3033892837563\\
1.75	0.301137432154804\\
1.76	0.298872405775953\\
1.77	0.296594754993816\\
1.78	0.294305029788325\\
1.79	0.292003779529141\\
1.8	0.289691552761483\\
1.81	0.287368896994028\\
1.82	0.285036358489007\\
1.83	0.28269448205458\\
1.84	0.280343810839621\\
1.85	0.277984886130996\\
1.86	0.275618247153457\\
1.87	0.273244430872216\\
1.88	0.270863971798338\\
1.89	0.268477401797002\\
1.9	0.266085249898755\\
1.91	0.263688042113818\\
1.92	0.261286301249553\\
1.93	0.258880546731149\\
1.94	0.25647129442562\\
1.95	0.254059056469189\\
1.96	0.251644341098117\\
1.97	0.249227652483066\\
1.98	0.246809490567043\\
1.99	0.244390350907\\
2	0.241970724519143\\
2.01	0.239551097728013\\
2.02	0.23713195201938\\
2.03	0.234713763897012\\
2.04	0.232297004743366\\
2.05	0.229882140684233\\
2.06	0.227469632457386\\
2.07	0.22505993528527\\
2.08	0.222653498751761\\
2.09	0.220250766683033\\
2.1	0.217852177032551\\
2.11	0.21545816177022\\
2.12	0.213069146775718\\
2.13	0.210685551736015\\
2.14	0.208307790047108\\
2.15	0.205936268719975\\
2.16	0.203571388290759\\
2.17	0.201213542735197\\
2.18	0.198863119387276\\
2.19	0.196520498862137\\
2.2	0.194186054983213\\
2.21	0.191860154713599\\
2.22	0.18954315809164\\
2.23	0.18723541817073\\
2.24	0.184937280963305\\
2.25	0.182649085389022\\
2.26	0.18037116322708\\
2.27	0.178103839072694\\
2.28	0.175847430297662\\
2.29	0.173602247015033\\
2.3	0.171368592047807\\
2.31	0.169146760901672\\
2.32	0.166937041741714\\
2.33	0.164739715373077\\
2.34	0.162555055225534\\
2.35	0.16038332734192\\
2.36	0.158224790370383\\
2.37	0.156079695560421\\
2.38	0.153948286762634\\
2.39	0.151830800432162\\
2.4	0.149727465635745\\
2.41	0.147638504062356\\
2.42	0.145564130037348\\
2.43	0.143504550540062\\
2.44	0.141459965224839\\
2.45	0.13943056644536\\
2.46	0.137416539282282\\
2.47	0.135418061574071\\
2.48	0.133435303951002\\
2.49	0.131468429872231\\
2.5	0.129517595665892\\
2.51	0.127582950572142\\
2.52	0.125664636789088\\
2.53	0.123762789521523\\
2.54	0.121877537032402\\
2.55	0.120009000696986\\
2.56	0.118157295059582\\
2.57	0.116322527892807\\
2.58	0.114504800259292\\
2.59	0.112704206575771\\
2.6	0.110920834679456\\
2.61	0.109154765896647\\
2.62	0.107406075113484\\
2.63	0.105674830848764\\
2.64	0.103961095328764\\
2.65	0.102264924563978\\
2.66	0.100586368427691\\
2.67	0.0989254707363237\\
2.68	0.0972822693314675\\
2.69	0.095656796163524\\
2.7	0.0940490773768869\\
2.71	0.0924591333965807\\
2.72	0.0908869790162829\\
2.73	0.089332623487655\\
2.74	0.0877960706109056\\
2.75	0.0862773188265115\\
2.76	0.0847763613080223\\
2.77	0.0832931860558745\\
2.78	0.0818277759921428\\
2.79	0.0803801090561542\\
2.8	0.0789501583008942\\
2.81	0.077537891990134\\
2.82	0.0761432736962073\\
2.83	0.0747662623983676\\
2.84	0.0734068125816569\\
2.85	0.0720648743362181\\
2.86	0.0707403934569834\\
2.87	0.0694333115436742\\
2.88	0.0681435661010446\\
2.89	0.0668710906393072\\
2.9	0.0656158147746766\\
2.91	0.0643776643299693\\
2.92	0.0631565614351987\\
2.93	0.0619524246281052\\
2.94	0.0607651689545648\\
2.95	0.0595947060688161\\
2.96	0.0584409443334515\\
2.97	0.0573037889191172\\
2.98	0.056183141903868\\
2.99	0.0550789023721257\\
3	0.0539909665131881\\
3.01	0.0529192277192403\\
3.02	0.0518635766828206\\
3.03	0.0508239014936912\\
3.04	0.0498000877350708\\
3.05	0.0487920185791828\\
3.06	0.047799574882077\\
3.07	0.0468226352776832\\
3.08	0.0458610762710549\\
3.09	0.0449147723307671\\
3.1	0.0439835959804272\\
3.11	0.0430674178892657\\
3.12	0.0421661069617703\\
3.13	0.0412795304263304\\
3.14	0.0404075539228603\\
3.15	0.0395500415893702\\
3.16	0.0387068561474556\\
3.17	0.0378778589866775\\
3.18	0.0370629102478065\\
3.19	0.0362618689049062\\
3.2	0.0354745928462314\\
3.21	0.0347009389539188\\
3.22	0.0339407631824492\\
3.23	0.0331939206358611\\
3.24	0.0324602656436974\\
3.25	0.0317396518356674\\
3.26	0.0310319322150083\\
3.27	0.0303369592305316\\
3.28	0.0296545848473413\\
3.29	0.0289846606162094\\
3.3	0.0283270377416012\\
3.31	0.0276815671483366\\
3.32	0.0270480995468818\\
3.33	0.0264264854972617\\
3.34	0.0258165754715877\\
3.35	0.0252182199151944\\
3.36	0.0246312693063825\\
3.37	0.024055574214763\\
3.38	0.0234909853582014\\
3.39	0.0229373536583607\\
3.4	0.0223945302948429\\
3.41	0.0218623667579294\\
3.42	0.0213407148999228\\
3.43	0.0208294269850922\\
3.44	0.0203283557382258\\
3.45	0.0198373543917953\\
3.46	0.019356276731737\\
3.47	0.0188849771418562\\
3.48	0.018423310646862\\
3.49	0.0179711329540396\\
3.5	0.0175283004935685\\
3.51	0.017094670457497\\
3.52	0.0166701008373811\\
3.53	0.0162544504606005\\
3.54	0.0158475790253608\\
3.55	0.0154493471343952\\
3.56	0.0150596163273774\\
3.57	0.01467824911206\\
3.58	0.0143051089941497\\
3.59	0.0139400605059358\\
3.6	0.0135829692336856\\
3.61	0.0132337018438214\\
3.62	0.0128921261078953\\
3.63	0.0125581109263782\\
3.64	0.012231526351278\\
3.65	0.0119122436076052\\
3.66	0.0116001351137026\\
3.67	0.0112950745004561\\
3.68	0.0109969366294056\\
3.69	0.0107055976097722\\
3.7	0.0104209348144226\\
3.71	0.0101428268947871\\
3.72	0.00987115379475114\\
3.73	0.00960579676353959\\
3.74	0.00934663836761228\\
3.75	0.00909356250159105\\
3.76	0.00884645439823723\\
3.77	0.00860520063749967\\
3.78	0.00836968915465303\\
3.79	0.00813980924754602\\
3.8	0.00791545158297997\\
3.81	0.00769650820223732\\
3.82	0.00748287252578056\\
3.83	0.00727443935714122\\
3.84	0.00707110488601945\\
3.85	0.00687276669061397\\
3.86	0.00667932373920262\\
3.87	0.00649067639099336\\
3.88	0.00630672639626593\\
3.89	0.00612737689582369\\
3.9	0.00595253241977585\\
3.91	0.00578209888566947\\
3.92	0.00561598359599097\\
3.93	0.00545409523505655\\
3.94	0.00529634386531102\\
3.95	0.00514264092305394\\
3.96	0.00499289921361238\\
3.97	0.00484703290597894\\
3.98	0.00470495752693398\\
3.99	0.00456658995467014\\
4	0.00443184841193801\\
};
\addplot [color=black,solid,forget plot,line width=1pt]
  table[row sep=crcr]{%
-1	0\\
-1	0.0539909665131881\\
};
\addplot [color=black,solid,forget plot,line width=1pt]
  table[row sep=crcr]{%
0	0\\
0	0.241970724519143\\
};
\addplot [color=black,solid,forget plot,line width=1pt]
  table[row sep=crcr]{%
1	0\\
1	0.398942280401433\\
};
\addplot [color=black,solid,forget plot,line width=1pt]
  table[row sep=crcr]{%
2	0\\
2	0.241970724519143\\
};
\addplot [color=black,solid,forget plot,line width=1pt]
  table[row sep=crcr]{%
3	0\\
3	0.0539909665131881\\
};

\node at (axis cs: -1.5,.2) {$(Z_0^b,Z_1^b)=$};
\node at (axis cs: -1.5,.1) {$(0,1)$};
\node at (axis cs: -.25,.1) {$(1,1)$};
\node at (axis cs: .5,.1) {$(0,0)$};
\node at (axis cs: 1.5,.1) {$(1,0)$};
\node at (axis cs: 2.25,.1) {$(0,1)$};
\node at (axis cs: 3.5,.1) {$(1,1)$};
\end{axis}
\end{tikzpicture}%

    \caption{The distribution of $\gamma(Z+\beta)$ with $\beta=\gamma=1$, $A=2$, and $N=2$. To find the distribution of $Z_0^b$ and $Z_1^b$, the interval $[-\ceil{\gamma A},\ceil{\gamma(A+2\beta)}]=[-2,4]$ is divided into sub-intervals $[k,k+1)$, which are labelled by their binary representation using \eqref{Zib} with $\hat{Z}$ as in \eqref{eq_zhat}. Then, $\mathcal{P}\{Z_i^b=1\}$ is the (normalized) sum of the areas under the curve for all sub-intervals with $Z_i^b=1$.}
   \label{fig_calculating noise dist}
\end{figure}
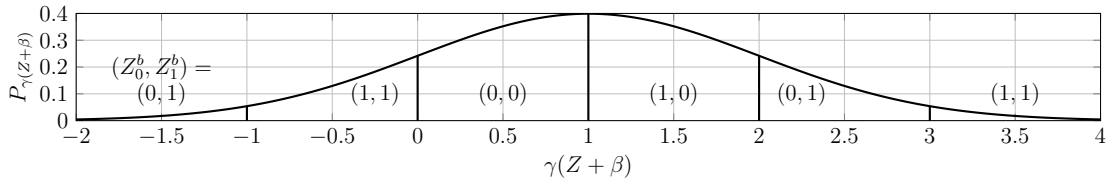

The distribution of $Z_i^b$ is a function of the bias $\beta$, the scaling factor $\gamma$, and the peak constraint $A$. Since it is assumed that $Z\in[-(A+\beta),A+\beta]$, then $\gamma(Z+\beta)\in[-\gamma A, \gamma(A+2\beta)]$. Thus, one can express the probability $\mathcal{P}\{Z_i^b=1\}$ as the integral of the probability density function of $\gamma(Z+\beta)$ over all intervals $[k,k+1)$ such that $k\in\{-\ceil{\gamma A},-\ceil{\gamma A}+1,\ldots,\ceil{\gamma( A+2\beta)}-1\}$, and such that $b_i^k$, the $i$-th bit in the binary representation of $k$, is equal to $1$. Thus
\begin{align}
\label{eqn_binnoise}
\alpha_i\triangleq\mathcal{P}\{Z_i^b=1\}=\frac{1}{T} \sum_{\substack{k=-\ceil{\gamma A}\\ b_i^k=1}}^{\ceil{\gamma( A+2\beta)}-1} Q\left(\frac{k-\gamma\beta}{\gamma\sigma}\right)-Q\left(\frac{k+1-\gamma\beta}{\gamma\sigma}\right),
\end{align}
%to ensure $\int_{-L}^{2^N-1}\frac{1}{T}P_{\gamma(Z+\beta)} dZ =1$
where $T=Q(\frac{-\ceil{\gamma A}-\gamma\beta}{\gamma\sigma})-Q(\frac{\ceil{\gamma( A+2\beta)}-\gamma\beta}{\gamma\sigma})$ is a normalization quantity to ensure that the integral of the density of $\gamma(Z+\beta)$ between $-\ceil{\gamma A}$ and $\ceil{\gamma( A+2\beta)}$ is equal to 1 when $\tilde{Y}\in[0,A+2\beta]$.\footnote{Note that compared to \cite{wfh_mlc} which uses a sum over the union of decision regions to characterize the binary channels for each level (bit-pipe), our approach is more direct since it is based on a binarization of noise.} Using similar representations as in Fig. \ref{fig_calculating noise dist}, one can conclude the following. If $\gamma\beta$ (mean of $\gamma(Z+\beta)$) is close to the midpoint of the quantized interval $[0,\gamma(A+2\beta)$, then all noise bits $Z_i^b$ follow a Bern$(0.5)$ distribution. Decreasing $\gamma\beta$ shifts the distribution of noise to the left relative to this midpoint, which moves $\alpha_i$ away from $0.5$ for some $i$, starting from the most-significant bit  and moving to bits of lower significance as $\gamma\beta$ decreases. In general, $\beta$ needs to be optimized to increase the achievable rates over the bit-pipes, but it is not useful to make $\beta$ large.

An example of $\alpha_i$ is given in Tab. \ref{fig_noisedistribfig} for $A=10$, $\beta=5$, and $\gamma=10$. Tab. \ref{fig_noisedistribfig} shows that although $\alpha_i$ improves in higher bit-pipes (deviates away from $0.5$), the improvement is gradual from bad bit-pipes (bit-pipes $0-3$) to useful but noisy bit-pipes (bit-pipe $4-6$) to noiseless ones (bit-pipe $7$). Thus, the bit-pipes can be modelled as BSC with cross-over probabilities in $[0,0.5]$ in general, contrary to \cite{adt_deterministic} where the cross-over probabilities are in $\{0,0.5\}$.

\begin{table}[t]
\renewcommand{\arraystretch}{1.3}
\centering
\begin{tabular}{c||c|c|c|c|c|c|c|c}
\hline
Bit-pipe &  $0$ & $1$ & $2$ & $3$ & $4$ & $5$ & $6$ & $7$ \\ \hline
$\alpha_i\triangleq\mathcal{P}\{Z_i^b =1\}$ & $0.5$ & $0.5$ & $0.5$ & $0.5$ & $0.54$ & $0.88$ & $0.08$ & $0$ \\\hline
%$\alpha_i$ & $0.5$ & $0.5$ & $0.5$ & $0.5$ & $0.46$ & $0.12$ & $0.08$ & $0$ \\\hline
\end{tabular}
\caption{Binary noise probability for a VBC with $A=10$, $\beta=5$, and $\gamma=10$.}
\label{fig_noisedistribfig}
\end{table}

While \eqref{eqn_binnoise} describes $Z_i^b$, it does not capture the joint distribution of $(Z_0^b,\ldots,Z_{N-1}^b)$ which are generally dependent (see Tab. \ref{tab_noise} for an example). To fully characterize the VBC, the conditional probabilities of noise are needed. In the following sections, $P_{Z_i^b|Z_{0}^b,\ldots,Z_{i-1}^b}$ for $i=1,\ldots,N-1$ will be needed. To calculate this, the joint probabilities $P_{Z_0^b,\ldots,Z_i^b}$ for $i=0,1,\ldots,N-1$ can be calculated using \eqref{eqn_binnoise} with $b_i^k=1$ adjusted as needed. Then the joint probabilities are used to find $P_{Z_i^b|Z_{0}^b,\ldots,Z_{i-1}^b}$ using Bayes' rule. For instance, for $N=4$, the probabilities $P_{Z_0^b}$, $P_{Z_0^b,Z_1^b}$, $P_{Z_0^b,Z_1^b,Z_2^b}$, and $P_{Z_0^b,Z_1^b,Z_2^b,Z_3^b}$, are calculated first, and then used to find $P_{Z_i^b|Z_{0}^b,\ldots,Z_{i-1}^b}$ for $i=1,\ldots,3$. 

The next section shows that the VBC decomposition is capacity-preserving, i.e., the VBC capacity can be made arbitrarily close to the IM/DD channel capacity by tuning $\beta$ and $\gamma$.

\begin{table}[t]
\renewcommand{\arraystretch}{1.3}
\centering
\begin{tabular}{c||c|c|c|c}
\hline
$(z_0^b,z_1^b)$                                &  $(0,0)$ & $(1,0)$ & $(0,1)$ & $(1,1)$  \\ \hline
$\mathcal{P}\{(Z_0^b, Z_1^b)=(z_0^b,z_1^b)\}$  & $0.1573$ & $0.1573$& $0.3426$ & $0.3426$ \\\hline
$\mathcal{P}\{Z_2^b =1 \mid Z_0^b, Z_1^b\}$    & $0.8640$ & $0.1360$& $0.0039$ & $0.0039$                      \\ \hline
\end{tabular}
\caption{Conditional noise probabilities for $A=2, \gamma=1, \beta=3$. In this example $\alpha_2=0.16$ differs from the conditionals $\mathcal{P}\{Z_2^b=1|Z_0^b,Z_1^b\}$, i.e., the variables are dependent.}
\label{tab_noise}
\end{table}

\subsection{The VBC model is Capacity-Preserving}
\label{sec_cappreserving}
The gap between $C^{\textnormal{VBC}_{\bar{\epsilon}}}$ and $C^{\text{IM/DD}}$ in the inequality in Lemma \ref{lem_capvbce} can be made arbitrarily small. This is discussed in the following theorem. 

\begin{theorem}\label{CapacityPreserving}
$C^{\text{VBC}_{\bar{\epsilon}}}$ can be made arbitrarily close to $C^{\text{IM/DD}}$ by optimizing $\gamma$ and $\beta$.
\end{theorem}
\begin{IEEEproof}
Consider an input distribution $P_X$ for the IM/DD channel, with mass points $(x_0,\ldots,x_K)\in[0,A]^K$ and masses $(a_0,\ldots,a_K)$. We focus on discrete distributions $P_X$ with a finite number of mass points since this achieves capacity \cite{smith_finitemasspoints,chk_discreteinputdist}. The achievable rate is then $I(X;Y)$.

The input $\boldsymbol{X}^b$ of the VBC generates $\hat{X}=\sum_{i=0}^{N-1}X_i^b2^i\in\hat{\mathcal{X}}$ where $N=\lceil\log_2\gamma(A+2\beta)\rceil$ as defined in \eqref{Eq:N} and $\hat{\mathcal{X}}=\{0,\frac{\gamma A}{2^N-1},\frac{2\gamma A}{2^N-1},\ldots,\frac{(2^N-1)\gamma A}{2^N-1}\}$. Note that for any $\eta>0$, there exists $N_0>0$ and $\hat{x}_k\in\hat{\mathcal{X}}$ for $k\in\{0,\ldots,K\}$, such that $|x_k-\frac{1}{\gamma}\hat{x}_k|<\eta$ $\forall k\in\{0,\ldots,K\}$ when $N>N_0$. This follows since increasing $N$ increases the resolution of the quantization of the interval $[0,A]$ into $2^N-1$ points. Thus, the distribution of $\frac{1}{\gamma}\hat{X}$ can be made arbitrarily close to $P_X$, and thus $I(X;Y)-I(\hat{X};Y)$ can be made arbitrarily small.

At the receiver, since the probability of truncation decreases to zero as $\beta$ increases, then for any $\eta>0$, there exists $\beta_0>0$ such that $I(\hat{X};Y)-I(\hat{X};Y')<\eta$ when $\beta>\beta_0$, where $Y'$ is the truncated $Y$ \eqref{eq_shifted_truncated}. Moreover, since $\hat{Y}=\gamma Y'$ is quantized into integers, the resulting quantization step for $Y'$ is $1/\gamma$, i.e., $\frac{1}{\gamma}(Y_0^b,\ldots,Y_{N-1}^b)$ is a quantization of $Y'$ with a quantization step of $\frac{1}{\gamma}$. From \cite[Theorem 8.3.1]{cover_elementsIT2}, it holds that $H(\frac{1}{\gamma}(Y_0^b,\ldots,Y_{N-1}^b))+\log(\frac{1}{\gamma})=H(Y_0^b,\ldots,Y_{N-1}^b)+\log(\frac{1}{\gamma})\to h(Y')$ as $\gamma$ increases, where $h(\cdot)$ and $H(\cdot)$ are the differential entropy and the entropy, respectively. Similarly, $H(Z_0^b,\ldots,Z_{N-1}^b)+\log(\frac{1}{\gamma})\to h(Z)$ as $\gamma$ increases. It follows that $I(\hat{X};Y_0^b,\ldots,Y_{N-1}^b)\to I(\hat{X};Y')$ as $\gamma$ increases. Thus, $I(\hat{X};Y_0^b,\ldots,Y_{N-1}^b)$ can be made arbitrarily close to $I(X;Y)$. Since $I(\hat{X};Y_0^b,\ldots,Y_{N-1}^b)$ is an achievable rate over VBC$_{\bar{\epsilon}}$ with $\beta$ large (so that $\bar{\epsilon}$ is negligible), this concludes the proof.
\end{IEEEproof}

\begin{table}[t]
\centering
\begin{tabular}{c|c|c|c|c|c||c}
\hline
$(\beta,\gamma)$ & $(5,0.2)$ & $(5,0.5)$ & $(5,1.5)$ & $(5,4)$ & $(5,6)$ & $(\infty,\infty)$, i.e., $\max_{P_X}I(X;Y)$ \\\hline
$A=100$ & $(3.4399,\ 6)$ & $(4.4305,\ 7)$ & $(4.6060,\ 8)$ & $(4.6226,\ 9)$ & $(4.6235,\ 10)$ & $(4.6531,\ \infty)$ \\\hline
$A=10$ & $(1.0000,\ 2)$ & $(1.6009,\ 4)$ & $(1.6042,\ 5)$ & $(1.7557,\ 7)$ & $(1.7568,\ 7)$ & $(1.7584,\ \infty)$ \\\hline
\end{tabular}
\caption{The numbers in the second and third rows show the pair $(\max_{P_{\boldsymbol{X}^b}}I(\boldsymbol{X}^b;\boldsymbol{Y}^b), N)$ for different values of $\beta$ and $\gamma$. The last column shows the IM/DD capacity (analogous to $N=\infty$). The capacity of the VBC approaches that of the IM/DD channel as $N$ grows.}
\label{Tab:CapPres}
\end{table}

Note that contrary to \cite[Thm. 1]{wfh_mlc} which refers to the capacity for a prior distribution $P_X$, Thm. \ref{CapacityPreserving} above refers to the capacity $C^{\rm IM/DD}=\max_{P_X} I(X;Y)$ as defined in~\eqref{eq_capimdd}. Table \ref{Tab:CapPres} shows that the values of $C^{\rm IM/DD}$ and $C^{{\rm VBC}_{\epsilon}}$ for different values $A$ and $N$ can be made arbitrarily close.

Coding over the VBC can be done in a manner similar to MLC/MSD and similar to `Hard-decision decoding at individual levels' \cite{wfh_mlc}, or somewhere in between with varying levels of complexity. The next sections discuss coding over the VBC model.

\section{IM/DD Channel Coding using the VBC}
\label{sec_codingschemes}
This section explains how a code for the VBC can be used in the IM/DD channel, and makes a connection with MLC/MSD. 

\subsection{From coding for the VBC to coding for the IM/DD channel}
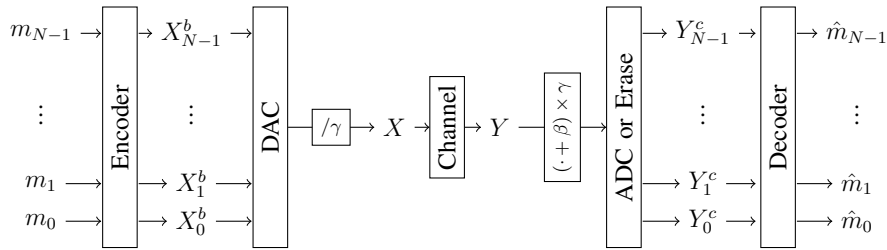
\begin{figure}
    \centering
    \begin{tikzpicture}[scale=1, every node/.style={scale=0.8}]
\node (mN-1) at (0,0) {$m_{N-1}$};
\node at (0,-1) {$\vdots$};
\node (m1) at (0,-2) {$m_{1}$};
\node (m0) at (0,-2.5) {$m_0$};

\node[rectangle,draw] (enc) at (1.05,-1.25) [minimum width = 4cm,rotate=90] {Encoder};
\draw[->] (mN-1) to ($(mN-1)+(.8,0)$);
\draw[->] (m1) to ($(m1)+(.8,0)$);
\draw[->] (m0) to ($(m0)+(.8,0)$);

\node (XN-1) at (2.0,0) {$X_{N-1}^b$};
\node at (2.0,-1) {$\vdots$};
\node (X1) at (2.0,-2) {$X_{1}^b$};
\node (X0) at (2.0,-2.5) {$X_0^b$};
\draw[->] ($(XN-1)-(.7,0)$) to (XN-1);
\draw[->] ($(X1)-(.7,0)$) to (X1);
\draw[->] ($(X0)-(.7,0)$) to (X0);

\node[rectangle,draw] (dac) at (3.05,-1.25) [minimum width= 4cm, rotate=90] {DAC};
\draw[->] (XN-1) to ($(XN-1)+(.8,0)$);
\draw[->] (X1) to ($(X1)+(.8,0)$);
\draw[->] (X0) to ($(X0)+(.8,0)$);

\node (X) at (4.7,-1.25) {$X$};
\draw[->] (dac) to node[sloped,draw,rectangle,fill=white] {\footnotesize $/\gamma$} (X);

\node[rectangle,draw] (ch) at (5.40,-1.25) [rotate=90]{Channel};
\draw[->] (X) to (ch);

\node (Y) at (6.1,-1.25) {$Y$};
\draw[->] (ch) to (Y);

\node[rectangle,draw] (adc) at (7.75,-1.25) [minimum width = 4cm,rotate=90] {ADC or Erase};
\draw[->] (Y) to node[sloped,draw,rectangle,fill=white,rotate=90] {\footnotesize $(\cdot+\beta)\times\gamma$} (adc);

\node (YN-1) at (8.8,0) {$Y_{N-1}^c$};
\node at (8.8,-1) {$\vdots$};
\node (Y1) at (8.8,-2) {$Y_{1}^c$};
\node (Y0) at (8.8,-2.5) {$Y_0^c$};

\draw[->] ($(YN-1)-(.8,0)$) to (YN-1);
\draw[->] ($(Y1)-(.8,0)$) to (Y1);
\draw[->] ($(Y0)-(.8,0)$) to (Y0);

\node[rectangle,draw] (enc) at (9.8,-1.25) [minimum width = 4cm, rotate=90] {Decoder};
\draw[->] (YN-1) to ($(YN-1)+(.7,0)$);
\draw[->] (Y1) to ($(Y1)+(.7,0)$);
\draw[->] (Y0) to ($(Y0)+(.7,0)$);

\node (mhN-1) at (10.85,0) {$\hat{m}_{N-1}$};
\node at (10.85,-1) {$\vdots$};
\node (mh1) at (10.85,-2) {$\hat{m}_{1}$};
\node (mh0) at (10.85,-2.5) {$\hat{m}_0$};

\draw[->] ($(mhN-1)-(.8,0)$) to (mhN-1);
\draw[->] ($(mh1)-(.8,0)$) to (mh1);
\draw[->] ($(mh0)-(.8,0)$) to (mh0);
\end{tikzpicture}
\caption{Coding for the IM/DD channel using the VBC. Messages $m_i$ with rate $R_i$ are encoded to binary symbols $X_i^b$, converted to decimal (DAC), scaled down by $\gamma$, and sent through the IM/DD channel. At the receiver, the received signal $Y$ is biased by $\beta$ and scaled by $\gamma$, erased (if $\gamma(Y+\beta)\notin[0,\gamma(A+2\beta)]$) or converted to binary (ADC) otherwise, and then used to decode~$\hat{m}_i$.}
\label{fig_imddchannelbp}
\end{figure}

Coding for the IM/DD channel can be realized by coding for the VBC and using the operations in Fig. \ref{fig_imddchannelbp}, some of which are similar to components that appear in \cite{newmultilevel_ih}. At the transmitter, the message $m$ with rate $R$ is split into multiple messages $m_i$, $i=0,\ldots,N-1$, with rates $R_i$ so that $R=\sum_iR_i$. The messages $m_i$ are encoded into binary codewords of length $n$ bits each, given by $(X_i^b(1),\ldots,X_i^b(n))$ for bit-pipe $i$. At each time instant, the encoded bits of all bit-pipes are mapped to an input symbol in $[0,A]$ using a mapper as follows. First, the transmitter generates the scaled signal $\gamma X(t)=\sum_i X_i^b(t)2^i$, $t=1,\ldots,n$ for some $N$ (as in \cite{newmultilevel_ih}). Then this is divided by $\gamma$ to produce an $X$ which satisfies the peak constraint, that is then sent over the channel. 

At the receiver, the corrupted symbol $Y$ is biased by $\beta$, scaled by $\gamma$, and erased if $\gamma(Y+\beta) \notin[0,\gamma(A+2\beta)]$. The result is then passed to a deterministic de-mapper which converts a decimal value into $N$ bits as in \eqref{eq_y bin}. This leads to received signals $(Y_i^c(1),\ldots,Y_i^c(n))$, $i=0,\ldots,N-1$, whose symbols are $0$, $1$, or $\xi$ (erasure) \eqref{eq_imdd_erasuress}. These signals are then decoded into $\hat{m}_i$.

The above description allows $(X_i^b(1),\ldots,X_i^b(n))$ to be generated jointly (dependently), which can be desired from an achievable rate point of view. The decoder can also use the output of all bit-pipes jointly to decode each $m_i$. The performance of this scheme can be arbitrarily close to capacity (cf. Thm. \ref{CapacityPreserving}). %A similar performance can be obtained using MSD as described next.

\vspace{-0.2cm}

\subsection{Multi-Level Coding (MLC) and Multi-Stage Decoding (MSD)}
In MLC, the transmitted signal constellation is labeled using bits, partitioned into subsets (commonly using set or block partitioning \cite{wfh_mlc}), and then each bit of the binary labels is encoded using a separate encoder. While our channel decomposition can be interpreted as a representation of the IM/DD channel in the form of binary channels, it can also be interpreted as a method of constructing $X$ at the transmitter in a manner that is similar to set partitioning. For instance, if $N=2$ and $\gamma=1$, then $X=\sum_{i=0}^{N-1}X_i^b2^i\in\{0,1,2,3\}$. The least significant bit $X_0^b$ defines the first level of partitioning which splits this set into $\{0,2\}$ or $\{1,3\}$ if $X_0^b=0$ or $1$, respectively, whereas the most significant bit $X_i^b$ defines the second level which splits each subset further to its individual components, similar to set partitioning \cite[Fig. 1]{wfh_mlc}.

At the receiver's side, after generating the transmitted signal as described above, an MSD receiver decodes from $Y$ directly. The receiver can also use $\boldsymbol{Y}^c$ to decode in a manner which is similar to MLC/MSD \cite{wfh_mlc}. It can decode $m_0$ from $(\boldsymbol{Y}^c(1),\ldots,\boldsymbol{Y}^c(n))$, followed by $m_i$, $i=1,\ldots,N-1$, successively in this order from $(\boldsymbol{Y}^c(1),\ldots,\boldsymbol{Y}^c(n))$ while using $m_0,\ldots,m_{i-1}$ as side information. The achievable rate is thus given by $I(m_0;\boldsymbol{Y}^c)+
\sum_{i=1}^{N-1} I(m_i;\boldsymbol{Y}^c|m_0,\ldots,m_{i-1})$. By the chain rule, this equals $I(m_0,\ldots,m_{N-1};\boldsymbol{Y}^c)=I(\boldsymbol{X}^b;\boldsymbol{Y}^c)$. This rate can be very close to the IM/DD channel capacity if $\boldsymbol{Y}^c$ is generated using large $\beta$ (no erasures) and large $\gamma$ (high resolution). However, to approach the optimal distribution of $X$, the bits $(X_0^b,\ldots,X_{N-1}^b)$ may need to be encoded jointly. Since independent encoding across the bit-pipes is more practical and more common in MLC/MSD, it will be used to compare the theoretical and practical (simulated) achievable rates of MLC and our proposed schemes in Sec. \ref{Sec:BCC} and \ref{sec_resultsanddiscussion}, respectively. %Since MLC/MSD with joint encoding and decoding achieves capacity, comparison with capacity will be considered instead.

Since one of the goals of the paper is to study the performance of BCC in the IM/DD channel, it is useful to simplify the encoding and decoding while benefiting from the input-output relation of bit-pipe $i$ described in Def. \ref{Def:VBC}. The next section considers independent encoding and proposes three decoding schemes, independent decoding (ID), state-assisted decoding (SD), and carry-over-assisted decoding (CD), in addition to implementations which can be realized using~BCC.

\section{Binary Channel Coding over the VBC}
\label{Sec:BCC}
This section presents simple schemes for the VBC with erasures, focusing on using BCC.

\vspace{-0.2cm}

\subsection{Independent Encoding}
\label{sec_encodingscheme}
Consider a subset $\mathcal{V}_{\boldsymbol{p}}$ of bit-pipes to be `on', i.e., where $0<p_i\leq1$ (since when $p_i=0$ then $X_i^b=0$ and the bit-pipe is `off').\footnote{Although information cannot be sent over bit-pipe $i$ when $p_i=1$, it is considered as `on' because it consumes from the available intensity $A$.} Then, the message $m$ with rate $R$ is split into $m_0, m_1,\dots,m_{N-1}$ with rates $R_0,\ldots,R_{N-1}$, respectively, where $R_i=0$ if $i\notin \mathcal V_{\boldsymbol{p}}$, whereas $R_i\geq 0$ if $i\in \mathcal V_{\boldsymbol{p}}$ as will be specified later. For $ i\in \mathcal V_{\boldsymbol{p}}$, $m_i$ is encoded into $(X_i^b(1),\ldots,X_i^b(n))$ using a binary code with i.i.d. Bern($p_i$) symbols and sent over the bit-pipe. For $i\notin\mathcal{V}_{\boldsymbol{p}}$, the transmitter sets  $X_i^b(t)=0$ for $t=1,\ldots,n$. Transmission follows the structure depicted in Fig. \ref{fig_imddchannelbp}, where at time $t$, the length $N$ vector $\boldsymbol{X}^b(t)=(X_0^b(t),\ldots,X_{N-1}^b(t))^T$ is converted to a decimal value using a DAC to obtain $\hat{X}(t)=\sum_{i=0}^{N-1}X_i^b(t)2^i$, scaled by $1/\gamma$, and then sent through the channel. This implies that $\hat{X}(t)$ is an integer, which ensures that $W_0^b(t)$ defined in \eqref{eq_cbin2} is zero.

This transmitter structure is common in MLC (cf. \cite{newmultilevel_ih}). A similar scheme was discussed in \cite[Sec. VII]{wfh_mlc}. However, the use of a DAC as a mapper and the independent encoding present a constraint in terms of shaping, since this is not necessarily capable of producing an input distribution of $X$ which coincides (or even closely resembles) the optimal input distribution of the IM/DD channel. This generally requires joint encoding over the bit-pipes and/or using a custom mapper or a distribution matcher \cite{fh_signalling,tw_A}. This issue will be revisited in Sec. \ref{sec_resultsanddiscussion}.

The choice of `on' bit-pipes and their respective $p_i$s needs to be optimized to maximize the achievable rate $\sum_{i=0}^{N-1}R_i$ while satisfying the constraints $X\in[0,A]$ and $\mathbb{E}[X]\leq E$. Thus, $\sum_{i\in \mathcal V_{\boldsymbol{p}}} 2^i\leq \gamma A$ and  $\sum_{i\in \mathcal V_{\boldsymbol{p}}} p_i2^{i}\leq \gamma E$. The achievable rate can be maximized once its expression is determined, which is the focus of the next subsections.

%\vspace{-0.5cm}
  
\subsection{Independent decoding (ID)}
\label{sec_independentcoding}
This scheme decodes over bit-pipes independently. For bit-pipe $i=0,\ldots,N-1$, the receiver decodes $m_i$ from $(Y_i^c(1),\ldots,Y_i^c(n))$ (instead of $(\boldsymbol{Y}^c(1),\ldots,\boldsymbol{Y}^c(n))$). This scheme is simple and can be easily realized in practice, since the bit-pipes are BSCs with erasures.

The receiver starts by decoding from the bit-pipe $i=0$, whose output is $(Y_0^c(1),\ldots,Y_0^c(n))$. Since encoding ensures that $\gamma X$ is an integer, this leads to $W_0^b(t)=0$ (cf. \eqref{eq_cbin2}). Hence this bit-pipe is a BSC with crossover probability $\alpha_0$ (BSC$(\alpha_0)$) described by $Y_0^b(t)=(X_0^b(t)+ Z_0^b(t))\bmod{2}$ concatenated with a BEC with erasure probability $\epsilon$ (BEC$(\epsilon)$). The receiver decodes $m_0$ from this bit-pipe's output. After decoding $m_0$, the receiver also recovers $Z_0^b(t)=(Y_0^b(t)+ X_0^b(t))\bmod{2}$ for all $t$ such that $Y_0^c(t)\neq\xi$. Then it computes $W_1^b(t)$ using \eqref{eq_cbin2}, and $\bar{Y}_i^c(t)= (Y_i^c(t)+W_i^b(t)) \mod 2= (X_1^b(t)+Z_1^b(t)) \mod 2$, $\forall t$ when $Y_i^c(t)\neq \xi$. Otherwise, $\bar{Y}_i^c(t)=\xi$. Again, this is a BSC$(\alpha_1)$ concatenated with a BEC$(\epsilon)$. The receiver decodes $m_1$ from $(\bar{Y}_i^c(1),\ldots,\bar{Y}_i^c(n))$. This procedure is repeated until all messages are decoded.\footnote{Note that the processing that is needed at the receiver to implement this successive decoding is all in binary (not real-valued operations). This feature can be appealing in practical implementations.} To present the achievable rate of this scheme, the following lemma is needed.

\begin{lemma}[{\cite{vw_channelwitherasure}}]
\label{lem_erasure}
Concatenating a discrete memoryless channel with capacity $C$ with an erasure channel that independently erases symbols with probability $\epsilon$ reduces capacity to $C(1-\epsilon)$.
\end{lemma}

This lemma holds if erasures are independent of the channel input. Otherwise, capacity may be higher than $C(1-\epsilon)$, and hence $C(1-\epsilon)$ becomes an achievable rate. The VBC belongs to the latter case. Now the achievable rate of the ID scheme can be expressed.

\begin{theorem}[Rate of ID]
\label{theorem_independentcodingachievablerate}
Given $\beta$ and $\gamma$, the achievable rate for the ID scheme is given by {$R_{\text{ID}}= (1-\bar{\epsilon})\max_{p_i}\sum_{i=0}^{N-1} (\mathsf{h}(p_i\circledast\alpha_i)-\mathsf{h}(\alpha_i))$}, 
%R_1(i) & = \max_{P_X} I(x_i^b;x_i^b) \\
%      & \leq 1-H(\alpha_i,1-\alpha_i)
where $\mathsf{h}(\cdot)$ is the binary entropy function, $\bar{\epsilon}$ is the erasure probability upper bound defined in \eqref{eq_perasure}, $\alpha_i$ is defined in \eqref{eqn_binnoise}, $p_i\circledast\alpha_i=p_i(1-\alpha_i)+(1-p_i)\alpha_i$, and the maximization is over the set of $p_i$ that satisfy $\sum_{i\in \mathcal V_{\boldsymbol{p}}}2^i\leq \gamma A$ and $\sum_{i\in \mathcal V_{\boldsymbol{p}}}p_i2^i\leq \gamma E$.
\end{theorem}
\begin{IEEEproof}
Bit-pipe $i$ is a BSC$(\alpha_i)$ concatenated with a (BEC$(\epsilon)$). Using Lemma \ref{lem_erasure}, the rate $R_i = (1-\epsilon)(\mathsf{h}(p_i\circledast\alpha_i)-\mathsf{h}(\alpha_i))$ is achievable over bit-pipe $i$. Since the erasure probability $\epsilon\leq\bar{\epsilon}$ \eqref{eq_perasure}, then $R_i\geq (1-\bar{\epsilon})(\mathsf{h}(p_i\circledast\alpha_i)-\mathsf{h}(\alpha_i))$. By summing over $i=0,\ldots,N-1$, and maximizing with respect to $p_i$ such that the input constraints are satisfied, the total rate $R_{\text{ID}}$ is obtained.
\end{IEEEproof}

\begin{remark}
The rate $R_{\rm ID}$ is achieved using a BSC and a BEC code. If the use of a BSC code only is desired, this can be realized by replacing erasures randomly with $0$ or $1$ at the receiver.
\end{remark}

\begin{figure}[t]
\centering
\begin{subfigure}[t]{0.47\textwidth}
         \centering
         \tikzset{every picture/.style={scale=.8}, every node/.style={scale=.8}}
%         \input{ID_vs_SNR}

% Minimal pgfplots version: 1.3
%  http://www.mathworks.com/matlabcentral/fileexchange/22022-matlab2tikz
\definecolor{mycolor1}{rgb}{0.38824,0.15686,0.15686}%
\begin{tikzpicture}

\begin{axis}[%
width=4.520833in,
height=3.458695in,
xmin=0,
xmax=20,
xlabel={$\frac{{A}}{\sigma}$ [dB]},
xmajorgrids,
ymin=0,
ymax=5,
ylabel={Rate (bits per transmission)},
ymajorgrids,
axis x line*=bottom,
axis y line*=left,
legend style={at={(axis cs: 0,5)},anchor=north west,legend cell align=left,align=left,draw=white!15!black}
]
\addplot [color=black,solid,line width=1pt]
  table[row sep=crcr]{%
0	0.160700000121326\\
3	0.484399999620577\\
6	0.938500000064862\\
9	1.52889999948447\\
12	2.26260000038404\\
15	3.10689999959583\\
18	4.01080000030609\\
21	4.9815000000622\\
24	5.95252679188549\\
27	6.93590799304662\\
30	7.92529397806425\\
};
\addlegendentry{Capacity};

\addplot[only marks,mark=x,mark options={},mark size=3pt,draw=blue,fill=blue] plot table[row sep=crcr,]{%
0	6.04709954474166e-05\\
.5  4.0539e-04\\
1   0.0021\\
1.5 0.0080\\
2   0.0225\\
2.5 0.0516\\
3	0.101600000656657\\
3.5 0.1646\\
4   .2597\\
4.5 0.3664\\
5   .4879\\
5.5 0.5994\\
6	0.727200000428726\\
6.5 .746\\
7   .903\\
7.5 .9434\\
8	.9805\\
8.5 .9943\\
9	1.00219999948542\\
9.5 1.1383\\
10  1.3248\\
10.5 1.4969\\
11  1.667\\
11.5 1.7153\\
12	1.91540000051423\\
12.5 1.9532\\
13  2.0839\\
13.5 2.2422\\
14  2.4314\\
14.5 2.6088\\
15	2.76170000064773\\
15.5 2.8181\\
16  2.9794\\
16.5  3.1515\\
17  3.3247\\
17.5  3.4969\\
18	3.66920000013168\\
18.5  3.7552\\
19  3.9290\\
19.5  4.0812\\
20  4.2669\\
21	4.62389999972758\\
22  4.8966\\
23  5.2420\\
24	5.60499999994849\\
27	6.60019999981467\\
30	7.57909999974305\\
};
\addlegendentry{$R_{\rm ID}$};

\addplot[only marks,mark=square,mark options={},mark size=3pt,draw=black,fill=black] plot table[row sep=crcr,]{%
     0    0.1607\\
%    1.0000    0.2398\\
    2.0000    0.3476\\
%    3.0000    0.4844\\
    4.0000    0.6405\\
%    5.0000    0.7929\\
    6.0000    0.9109\\
%    7.0000    1.0352\\
    8.0000    1.2722\\
%    9.0000    1.5146\\
   10.0000    1.7360\\
%   11.0000    1.9547\\
   12.0000    2.2406\\
%   13.0000    2.5213\\
   14.0000    2.7615\\
%   15.0000    3.0579\\
   16.0000    3.3961\\
%   17.0000    3.6906\\
   18.0000    4.0105\\
%   19.0000    4.3285\\
   20.0000    4.6423\\
   };
\addlegendentry{$R_{{\rm MLC}}$};

\addplot [color=red,solid]
  table[row sep=crcr]{%
0	0.0413\\
2.5	0.1232\\
5	0.3341\\
7.5	0.7589\\
10	1.3926\\
12.5	2.1477\\
15	2.9526\\
17.5	3.7748\\
20	4.6026\\
};
\addlegendentry{Uniform dist. \cite[(18)]{lmw_bounds}};

\addplot [color=red,dashed]
  table[row sep=crcr]{%
0     0\\
     2     1\\
     4     2\\
     6     2\\
     8     3\\
    10     4\\
    12     4\\
    14     5\\
    16     6\\
    18     6\\
    20     7\\
};    
\addlegendentry{Approximation in \cite{adt_deterministic}};
    
\end{axis}
\end{tikzpicture}%

         \caption{Peak constraint only ($E=A$).}
		\label{capfig}
     \end{subfigure}
     \hspace{.5cm}
     \begin{subfigure}[t]{0.47\textwidth}
         \centering
		\tikzset{every picture/.style={scale=.8}, every node/.style={scale=.8}}
%		\input{ID_vs_SNR_rho_1by3}

% Minimal pgfplots version: 1.3
%  http://www.mathworks.com/matlabcentral/fileexchange/22022-matlab2tikz
\definecolor{mycolor1}{rgb}{0.46600,0.67400,0.18800}%
\begin{tikzpicture}

\begin{axis}[%
width=4.520833in,
height=3.455223in,
xmin=0,
xmax=20,
xlabel={$\frac{{A}}{\sigma}$ [dB]},
xmajorgrids,
ymin=0,
ymax=5,
ylabel={Rate (bits per transmission)},
ymajorgrids,
axis x line*=bottom,
axis y line*=left,
legend style={at={(axis cs:0,5)},anchor=north west,legend cell align=left,align=left,draw=white!15!black}
]

\addplot [color=black,solid,line width=1pt]
  table[row sep=crcr]{%
0	0.1443\\
2.5	0.3739\\
5	0.7251\\
7.5	1.1170\\
10	1.6222\\
12.5	2.2267\\
15	2.9025\\
17.5	3.6467\\
20	4.3891\\
%22.5	5.1722\\
%24  5.5303\\
%25	5.7148\\
%27.5	4.24205056155478\\
%30	3.70044164190238\\
};
\addlegendentry{Capacity};

\addplot[only marks,mark=square,mark options={},mark size=3pt,draw=black,fill=black] plot table[row sep=crcr,]{%
     0    0.1369\\
    2.0000    0.2993\\
    4.0000    0.5582\\
    6	0.8006\\
    8.0000    1.1227\\
   10.0000    1.5425\\
12	2.0598\\
   14.0000    2.5361\\
16  3.1494\\
18	3.7601\\
20  4.3614\\
};
\addlegendentry{$R_{\rm MLC}$};

\addplot[only marks,mark=x,mark options={},mark size=3pt,draw=blue,fill=blue] plot table[row sep=crcr,]{%
     0    0.0001\\
    0.5000    0.0001\\
    1.0000    0.0005\\
    1.5000    0.0038\\
    2.0000    0.0193\\
    2.5000    0.0439\\
3	0.084544787\\
3.5000    0.1392\\
    4.0000    0.2166\\
    4.5000    0.3107\\
    5.0000    0.4112\\
    5.5000    0.5250\\
    6	0.630591264\\
6.5000    0.7169\\
    7.0000    0.7576\\ 
    7.5000    0.7983\\
    8.0000    0.8606\\
    8.5000    0.9286\\
    9.0000    0.9559\\
    9.5000    0.9774\\
   10.0000    1.1538\\
   10.5000    1.3152\\
   11.0000    1.4727\\
   11.5000    1.6004\\      
12	1.7533\\
12.5000    1.7875\\
   13.0000    1.8487\\
   13.5000    2.0459\\
   14.0000    2.1253\\
   14.5000    2.3773\\
15	2.512960555\\
15.5  2.6370\\
16  2.7281\\
16.5  2.8793\\
17  3.0626\\
17.5  3.2310\\
18	3.4099\\
18.5  3.5477\\
19  3.6332\\
19.5  3.8164\\
20  3.9665\\
21	4.356233443\\
22 4.6308\\
24	5.326852529\\
27	6.305557138\\
30	7.283524668\\
};
\addlegendentry{$R_{\rm ID}$};

\addplot [color=red,solid]
  table[row sep=crcr]{%
  0 0.028886276316062503430306522980557\\
3  0.10871251967552568034850964302105\\
  6  0.36016005610185135704262728423362\\
  9  0.91956324008468704103322246757252\\
 12   1.7467556016003180411989797241752\\
 15    2.693706929352218677719012421044\\
 18   3.6772639562859068470616821992874\\
 20   4.3389917277966257365506238357455\\
};
\addlegendentry{Truncated Exp. dist. \cite[(10)]{lmw_bounds}};

\end{axis}
\end{tikzpicture}%

         \caption{Peak and average constraints, $E=A/3$.}
		\label{fig_avrgconst}
     \end{subfigure}
     \caption{Achievable rates for ID ($R_{\text{ID}}$) compared to capacity.}
     \label{capfigfinal}
\end{figure}
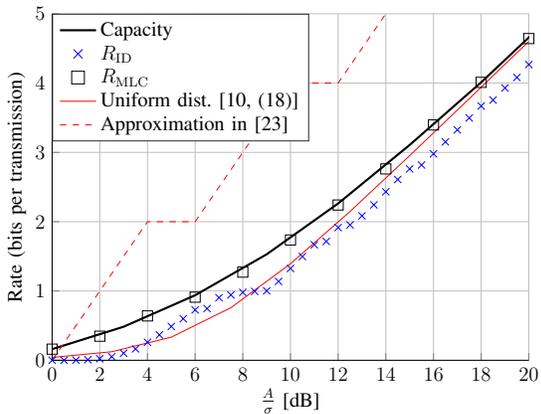
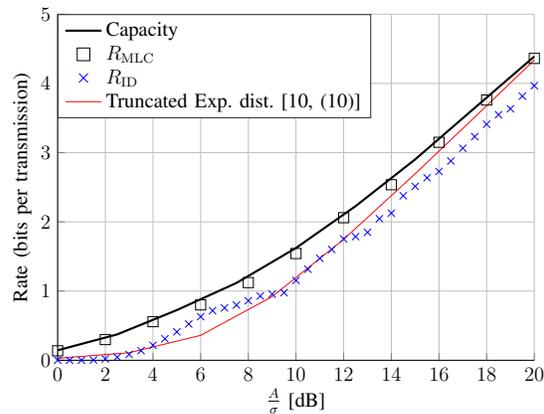

The achievable rate $R_{\rm ID}$ (with $\beta$, $\gamma$, and $p_i$ optimized using uniform sampling)\footnote{Choosing $\gamma=a$ and $\gamma=2a$ for some $a$ leads to the same performance because $\gamma=2a$ results in a shifted binary representation of $\hat{X}$, $\hat{Y}$, and $\hat{Z}$ by 1 bit compared to $\gamma=a$. Thus, it suffices to search over $\gamma\in[a,2a)$ for some $a$. However, $a$ should be chosen so that $N$ is larger than capacity, since otherwise the achievable rate of the ID scheme cannot approach capacity due to insufficient bit-pipes.} is shown in Fig. \ref{capfigfinal}, along with the channel capacity evaluated using numerical optimization \cite{chk_discreteinputdist}, the achievable rate using a continuous input distribution (uniform and truncated-exponential) from \cite{lmw_bounds}, and the achievable rate using MLC/MSD under independent encoding (with optimized $p_i$). The approximation obtained using the approach in \cite{adt_deterministic}, which is given by $\left\lceil\frac{1}{2}\log_2(\frac{A^2}{\sigma^2})\right\rceil$ is also shown for comparison. Note that $R_{\rm ID}$ follows a similar trend to capacity as a function of SNR $\left(\frac{A}{\sigma}\right)$ within a gap. Since the ID scheme is a simplification of MSD where decoding is performed on binary outputs of bit-pipes independently (hard-decision decoding), its performance is lower than MLC/MSD (which uses soft-decision decoding) in return for lower complexity. The plateaus and transitions in $R_{\rm ID}$ occur when a new `good' bit-pipe emerges among the set of bit-pipes BSC$(\alpha_i)$, $i=0,\ldots,N-1$ (cf. Tab. \ref{Fig:PzTransitions}). 

\begin{table}
\centering
\begin{tabular}{c||c|c|c|c}
bit-pipe $i$ & $A/\sigma=9$ dB & $A/\sigma=9.1$ dB & $A/\sigma=9.2$ dB & $A/\sigma=9.3$ dB \\\hline
$0,\ldots,7$ & $0.5$		   & $0.5$			   & $0.5$			   & $0.5$    \\\hline
$8$ 		 & $0.4825$		   & $0.4599$		   & $0.4641$		   & $0.4544$ \\\hline
$9$ 		 & $0.4755$		   & $0.2395$		   & $0.1988$		   & $0.2046$ \\\hline
$10$		 & $0.0001$		   & $0.0296$		   & $0.0396$		   & $0.0303$ \\\hline
\end{tabular}
\caption{Value of $\alpha_i'=\min\{\alpha_i,1-\alpha_i\}$ under a peak constraint only as $A/\sigma$ increases after a `plateau' of $R_{\rm ID}$. At $A/\sigma=9$ dB, only bit-pipe $10$ is good. At $A/\sigma=9.1$ dB, bit-pipe $9$ emerges as a `good' bit-pipe. Note that $\alpha_i'$ is used here to show channel improvement as a decreasing behaviour of $\alpha_i'$ bearing in mind that a BSC$(\alpha_i)$ and a BSC$(\alpha_i')$ have the same capacity.}
\label{Fig:PzTransitions}
\end{table}

The gap between $R_{\rm ID}$ and capacity is attributed to the following reasons. First, the use of independent encoding constrains shaping (to be discussed in Sec. \ref{sec_resultsanddiscussion}). Second, this scheme neglects information inherent in the previous bit-pipes' decoding outcomes ($m_0,\ldots,m_{i-1}$) when decoding $m_i$. Nonetheless, this scheme is fairly simple and can be realized using BCC (an implementation using polar codes is presented in Sec. \ref{sec_polarindependent}). To close this gap, we consider next a decoder which benefits from side information available at the receiver.

\subsection{State-Assisted Decoding (SD)}
\label{sec_modfullstate1}
A scheme based on the theory of the channel with a state known at the receiver is proposed first in this subsection. Then it is simplified to enable using standard BCCs.

\subsubsection{The SD$_q$ Scheme -- A Channel with State Perspective}
After decoding $m_0$, the receiver can use it as side information when decoding $m_{1}$. To convert the knowledge of $m_0$ into a property of the BSC over bit-pipe $1$, the receiver can use $m_0$ and $Y_0^c(t)$ to obtain the noise $Z_0^b(t)$, for all $t$ where $Y_0^c(t)\neq \xi$. Then it can use $Z_0^b(t)$ as side-information while decoding $m_1$ from $(Y_1^c(1),\ldots,Y_1^c(n))$. The dependence of $Z_1^b(t)$ on $Z_{0}^b(t)$ (cf. Tab. \ref{tab_noise}) makes $Z_{0}^b(t)$ useful when decoding $m_1$. Thus, bit-pipe $1$ becomes a BSC with state $Z_0^b(t)$ known at the decoder \cite[Ch. 7]{ElgamalKim} (with transition probability $P_{Y_1^b|X_1^b,Z_0^b}$). Its capacity can be expressed using the following lemma.

\begin{lemma}[\cite{ElgamalKim}]
\label{lem_state}
The capacity of a channel with transition probability $P_{Y|X,S}$, where $X$ is the input, $Y$ is the output, and $S$ is the state known only at the decoder, is $C= \max_{P_X}\;I(X;Y|S)$.
\end{lemma}

This capacity can be achieved by treating $(Y,S)$ as the channel output, and is in general larger than the capacity of the channel described by $P_{Y|X}=\mathbb{E}_S[P_{Y|X,S}]$. This can be exploited to improve upon the rate of the ID scheme as follows. After decoding $m_0$ and calculating $W_1^b(t)$ for all $t$ with $Y_0^c(t)\neq \xi$ as in the ID scheme, the receiver subtracts $W_1^b(t)$ from $Y_1^c(t)$ to obtain 
\begin{align}\label{Yibar}
\bar{Y}_1^c(t)=\begin{cases}
(Y_1^b(t)+W_1^b(t))\bmod 2=(X_1^b(t)+Z_1^b(t))\bmod 2, & \text{ if } Y_1^c(t)\neq \xi\\
\bar{Y}_1^c(t)=\xi, &\text{otherwise.}
\end{cases}
\end{align}
Combining this with $Z_0^b(t)$, the resulting channel can be interpreted as a channel with state $Z_0^b(t)$ known at the decoder, concatenated with a BEC$(\epsilon)$. The decoder decodes $m_1$ from the output of this channel. The same is repeated to decode $m_2$ from $(\bar{Y}_2^c(t),Z_0^b(t),Z_1^b(t))$. This repeats until decoding $m_{N-1}$ from $(\bar{Y}_{N-1}^c(t),Z_{0}^b(t),\ldots,Z_{N-2}^b(t))$.

Note that this bears some similarity with MLC/MSD, except that MLC/MSD uses side information from $(Z_{0}^b,\ldots,Z_{i-1}^b)$ (obtained from $(Y_{i-1}^c,\ldots,Y_{0}^c)$ and $(m_{i-1},\ldots,m_0)$) and also from $(Y_{i+1}^c,\ldots,Y_{N-1}^c)$ when decoding $m_i$ from $Y_i^c$, whereas this scheme only uses $(Z_{0}^b,\ldots,Z_{i-1}^b)$. This makes the complexity of the SD scheme between that of the ID and the MLC/MSD schemes, and similarly for the performance as we shall see in Fig. \ref{capfigAllSchemes}.

The number of inputs to the decoder at different bit-pipes varies: For bit-pipe $i$, the decoder uses $i+1$ input streams to decode the message ($1$ for the channel output $\bar{Y}_{i}^c(t)$ and $i$ for the state $(Z_{0}^b(t),\ldots,Z_{i-1}^b(t))$. It is better to unify the decoding over all bit-pipes to have the same number of inputs at the decoder. Moreover, the decoder of a bit-pipe uses the noise from all lower bit-pipes as a state, although the noise on some of these lower bit-pipes may only contribute marginal rate improvement. Thus, decoding can be simplified by reducing the number of inputs if possible. Using these observations, the scheme can be  generalized as follows.

Let $Z_{j}^b(t)=0$ for all $j<0$ and for all $t$. Also, denote by $q$ the number of state bits, and by
\begin{align}
\label{A_i}
\mathcal{A}_i=\{a_{i1},\ldots,a_{iq}\}\subseteq\{i-N+1,\ldots,i-1\}
\end{align}
the set of bit-pipes whose noise will be used as a state at the decoder of bit-pipe $i$. Then a generic form of the SD scheme, to be denoted the SD$_q$ scheme, can be described as follows: The message $m_i$ is decoded from $(\bar{Y}_i^c(1),\ldots,\bar{Y}_i^c(n))$ and $(\boldsymbol{S}_i^c(1),\ldots,\boldsymbol{S}_i^c(n))$, where 
\begin{align}
\label{S_ic}
\boldsymbol{S}_i^c(t)=(Z_{a_{i1}}^b(t),\ldots,Z_{a_{iq}}^b(t))\triangleq \boldsymbol{S}_i^b(t)
\end{align}
if $Y_i^c(t)\neq\xi$ and $\boldsymbol{S}_i^c(t)=\xi$ otherwise. In this case, the decoder uses $q+1$ input streams to decode $m_i$ for all $i$, and the $q$ bit-pipes whose noise will be used in the decoding process can be optimized by choosing $\mathcal{A}_i$. The most general case is $q=N-1$ where noise on all previous bit-pipes is used as a state. But $q$ can be chosen to be less than $N-1$ to simplify the scheme.

Knowing $\mathcal{P}\{Z_i^b=1|Z_{a_{i1}}^b,\ldots,Z_{a_{iq}}^b\}$, the achievable rate of the SD$_q$ scheme is expressed next.

\begin{theorem}[Rate of SD$_q$]
\label{theorem_stateawarecodingachievablerate}
For a given $q\in\{1,\ldots,N-1\}$, and given $\beta$, $\gamma$, and $\mathcal{A}_i$ as defined in \eqref{A_i}, the SD$_q$ scheme achieves the rate $R_{\text{SD}_q}= (1-\bar{\epsilon})\max_{p_i}\sum_{i=0}^{N-1} \mathbb{E}_{\boldsymbol{S}_i^b}[\mathsf{h}(p_i\circledast \alpha_i(\boldsymbol{S}_i^b))-\mathsf{h}(\alpha_i(\boldsymbol{S}_i^b))]$, where $\boldsymbol{S}_i^b$ is defined in \eqref{S_ic}, $\mathbb{E}_{\boldsymbol{S}_i^b}$ is the expectation with respect to $\boldsymbol{S}_i^b$, $\alpha_i(\boldsymbol{S}_i^b)=\mathcal{P}\{Z_i^b=1|\boldsymbol{S}_i^b\}$ (cf. Sec. \ref{sec_noise}), and the maximization is subject to $\sum_{i\in \mathcal V_{\boldsymbol{p}}}2^i\leq \gamma A$ and $\sum_{i\in \mathcal V_{\boldsymbol{p}}}p_i2^i\leq \gamma E$.
\end{theorem}
\begin{IEEEproof}
Bit-pipe $i$ is a BSC with a state $\boldsymbol{S}_i^b$ known at the decoder, concatenated with a BEC$(\epsilon)$. The crossover probability of the BSC depends on $\boldsymbol{S}_i^b$, and hence is denoted $\alpha_i(\boldsymbol{S}_i^b)$ which can be calculated as described in Sec. \ref{sec_noise}. Using Lemma \ref{lem_state}, the achievable rate over bit-pipe $i$ is $R_i=(1-\epsilon)I(X_i^b;Y_i^b|\boldsymbol{S}_i^b)$ (since $\bar{Y}_i^c=Y_i^b$ when $Y_i^c\neq \xi$). But, $R_i=(1-\epsilon)I(X_i^b;{Y}_i^b|\boldsymbol{S}_i^b)=(1-\epsilon)\mathbb{E}_{\boldsymbol{S}_i^b}[\mathsf{h}(p_1\circledast\alpha_i(\boldsymbol{S}_i^b))-\mathsf{h}(\alpha_i(\boldsymbol{S}_i^b))]\geq (1-\bar{\epsilon})\mathbb{E}_{\boldsymbol{S}_i^b}[\mathsf{h}(p_1\circledast\alpha_i(S_i^b))-\mathsf{h}(\alpha_i(\boldsymbol{S}_i^b))]$. Summing over all bit-pipes and maximizing with respect to $p_i$ proves the theorem.
\end{IEEEproof}

Note that $R_{{\rm SD}_{q}}\geq R_{\rm ID}$ since $\mathbb{E}_{\boldsymbol{S}_i^b}[I(X_i^b;Y_i^b|\boldsymbol{S}_i^b)]\geq I(X_i^b;Y_i^b)$ by the convexity of the mutual information in the transition probability for given input distribution \cite{cover_elementsIT2}. Thus, the SD$_q$ scheme outperforms the ID scheme. Fig. \ref{capfigAllSchemes} shows that this scheme approaches capacity at moderate and high SNR. This scheme is in fact closer to MLC/MSD than ID since it exploits information inherent in other bit-pipes' output while decoding the output of bit-pipe $i$. However, it differs from MLC/MSD in that it only exploits lower bit-pipes only.

The achievable rate $R_{{\rm SD}_q}$ can be maximized by choosing $\beta$, $\gamma$, and $\mathcal{A}_i$ for a given $q$. In terms of implementation, this scheme requires using a code for a channel with a binary input and $q+1$ binary outputs. Since the main aim of this paper is to investigate the limits of using BCC, a simple implementation of the SD$_q$ scheme is proposed next.

\subsubsection{The SD-BSC Scheme}
To implement an SD scheme using BCC, the state $\boldsymbol{S}_i^b$ should be used in a way that improves the BSC from $X_i^b$ to $Y_i^b$, i.e., decreases its crossover probability. To this end, consider the following toy example, focusing on the probabilities given in Tab. \ref{tab_noise}. In this example, $\alpha_i=\mathcal{P}\{Z_2^b=1\}=0.16$, i.e., bit-pipe $2$ is a BSC$(0.16)$ under the ID scheme. By examining Tab. \ref{tab_noise}, it can be seen that when $(Z_0^b,Z_1^b)=(0,0)$, then bit-pipe $2$ is a BSC$(0.864)$ since $\mathcal{P}\{Z_2^b=1|(Z_0^b,Z_1^b)=(0,0)\}=0.864$, i.e., bit-pipe $2$ flips its input bit most of the time. Otherwise, when $(Z_0^b,Z_1^b)\neq (0,0)$, bit-pipe $2$ is a BSC with crossover probabilities $0.136$, $0.0039$, and $0.0039$ when $(Z_0^b,Z_1^b)$ equals $(1,0)$, $(0,1)$, and $(1,1)$, respectively. If the receiver knows $(Z_0^b,Z_1^b)$, it can use it to improve this channel by flipping the output $Y_2^b$ when $(Z_0^b,Z_1^b)=(0,0)$. This converts bit-pipe $2$ into a BSC$(0.136)$ when $(Z_0^b,Z_1^b)=(0,0)$. By averaging the cross-over probabilities over $(Z_0^b,Z_1^b)$, bit-pipe $2$ becomes a BSC$(0.0455)$. This amounts to a rate improvement from $0.3657$ bits to $0.733$ over bit-pipe $2$.

This idea can be used to design the SD-BSC scheme, where the channel state is used to improve the BSC modelling each bit-pipe. Given $\mathcal{A}_i$ as defined in \eqref{A_i} the decoder of bit-pipe $i$ receives $(\overline{Y}_i^c(1),\ldots,\overline{Y}_i^c(n))$ from the output of the bit-pipe and $(\boldsymbol{S}_i^c(1),\ldots,\boldsymbol{S}_i^c(n))$ (defined in \eqref{S_ic}) from the decoders of the lower bit-pipes. It also knows $\mathcal{P}\{Z_i^b=1|\boldsymbol{S}_i^c\}$. Then it constructs 
\begin{align}\label{y_breve_c}
\breve{Y}_i^c(t)=\begin{cases}
\overline{Y}_i^b(t) & \text{if } \overline{Y}_i^c(t)\neq \xi \text{ and } \mathcal{P}\{Z_i^b=1|\boldsymbol{S}_i^c=\boldsymbol{S}_i^c(t)\}\leq 0.5\\
(1+\overline{Y}_i^b(t))\bmod 2 & \text{if } \overline{Y}_i^c(t)\neq \xi \text{ and } \mathcal{P}\{Z_i^b=1|\boldsymbol{S}_i^c=\boldsymbol{S}_i^c(t)\}>0.5\\
\xi & \text{if } \overline{Y}_i^c(t)= \xi.
\end{cases}
\end{align}
The resulting channel with input $X_i^b$ and output $\breve{Y}_i^c$ is a BSC concatenated with an erasure channel. The achievable rate of this scheme, denoted th SD-BSC$_q$ scheme, is given next.

\begin{theorem}[Rate of SD-BSC$_q$]
\label{theorem_SD_BSC}
For a given $q\in\{1,\ldots,N-1\}$, and given $\beta$, $\gamma$, and $\mathcal{A}_i$ as defined in \eqref{A_i}, the SD-BSC$_q$ scheme achieves the rate $R_{\text{SD-BSC}_q}= (1-\bar{\epsilon})\max_{p_i}\sum_{i=0}^{N-1} \mathsf{h}(p_i\circledast \breve{\alpha}_i)-\mathsf{h}(\breve{\alpha}_i)$, where $\breve{\alpha}_i
=\mathbb{E}_{\boldsymbol{S}_i^b}\left[
\min\left\{
\mathcal{P}\{Z_i^b=1|\boldsymbol{S}_i^b\},1-\mathcal{P}\{Z_i^b=1|\boldsymbol{S}_i^b\}\right\}\right]$, $\boldsymbol{S}_i^b$ is defined in \eqref{S_ic}, and $p_i$ satisfies $\sum_{i\in \mathcal V_{\boldsymbol{p}}}2^i\leq \gamma A$ and $\sum_{i\in \mathcal V_{\boldsymbol{p}}}p_i2^i\leq \gamma E$.
\end{theorem}
\begin{IEEEproof}
From the definition of the output of the BSC in \eqref{y_breve_c}, this BSC has crossover probability $\mathcal{P}\{Z_i^b=1|\boldsymbol{S}_i^b\}$ if $\mathcal{P}\{Z_i^b=1|\boldsymbol{S}_i^b\}\leq0.5$ and has crossover probability $1-\mathcal{P}\{Z_i^b=1|\boldsymbol{S}_i^b\}$ otherwise, given $\boldsymbol{S}_i^b$. This can be written as $\min\left\{
\mathcal{P}\{Z_i^b=1|\boldsymbol{S}_i^b\},1-\mathcal{P}\{Z_i^b=1|\boldsymbol{S}_i^b\}\right\}$ for a given $\boldsymbol{S}_i^b$. The crossover probability of the channel when considering all possible realizations of $\boldsymbol{S}_i^b$ is thus the expectation in $\breve{\alpha}_i$. Summing the achievable rates over all bit-pipes and considering the erasure probability upper bound $\bar{\epsilon}$ yields $R_{{\rm SD-BSC}_q}$.
\end{IEEEproof}

\begin{figure}
    \centering
\begin{subfigure}[t]{0.47\textwidth}
         \centering
         \tikzset{every picture/.style={scale=.8}, every node/.style={scale=.8}}
%         \input{rates_vs_SNR_peak_only}%
%         \input{rates_vs_SNR_peak_only_with_SD_BSC}

% Minimal pgfplots version: 1.3
%  http://www.mathworks.com/matlabcentral/fileexchange/22022-matlab2tikz
\definecolor{mycolor1}{rgb}{0.38824,0.15686,0.15686}%
\begin{tikzpicture}

\begin{axis}[%
width=4.520833in,
height=3.458695in,
xmin=0,
xmax=20,
xlabel={$\frac{{A}}{\sigma}$ [dB]},
xmajorgrids,
ymin=0,
ymax=5,
ylabel={Rate (bits per transmission)},
ymajorgrids,
axis x line*=bottom,
axis y line*=left,
legend style={at={(axis cs: 0,5)},anchor=north west,legend cell align=left,align=left,draw=white!15!black}
]

\addplot [color=black,solid,line width=1pt]
  table[row sep=crcr]{%
0	0.160700000121326\\
3	0.484399999620577\\
6	0.938500000064862\\
9	1.52889999948447\\
12	2.26260000038404\\
15	3.10689999959583\\
18	4.01080000030609\\
21	4.9815000000622\\
24	5.95252679188549\\
27	6.93590799304662\\
30	7.92529397806425\\
};
\addlegendentry{Capacity};

\addplot[only marks,mark=x,mark options={},mark size=3pt,draw=blue,fill=blue] plot table[row sep=crcr,]{%
0	6.04709954474166e-05\\
%.5  4.0539e-04\\
1   0.0021\\
%1.5 0.0080\\
2   0.0225\\
%2.5 0.0516\\
3	0.101600000656657\\
%3.5 0.1646\\
4   .2597\\
%4.5 0.3664\\
5   .4879\\
%5.5 0.5994\\
6	0.727200000428726\\
%6.5 .746\\
7   .903\\
%7.5 .9434\\
8	.9813\\
%8.5 .9943\\
9	1.00219999948542\\
%9.5 1.1383\\
10  1.3248\\
%10.5 1.4969\\
11  1.667\\
%11.5 1.7153\\
12	1.91540000051423\\
%12.5 1.9532\\
13  2.0839\\
%13.5 2.2422\\
14  2.4314\\
%14.5 2.6088\\
15	2.76170000064773\\
%15.5 2.8181\\
16  2.9794\\
%16.5  3.1515\\
17  3.3247\\
%17.5  3.4969\\
18	3.66920000013168\\
%18.5  3.7552\\
19  3.9290\\
%19.5  4.0812\\
20  4.2669\\
};
\addlegendentry{$R_{\rm ID}$};

\addplot[only marks,mark=o,mark options={},mark size=3pt,draw=red,fill=red] plot table[row sep=crcr,]{%
0	7.4655e-05\\
1	0.0027\\
2	0.0280\\
3	0.1251\\
4   0.3261\\
5   0.5916\\
6	0.8135\\
7   0.9503\\
8	1.025\\
9	1.3480\\
10  1.6450\\
11  1.8555\\
12	2.1194\\
13  2.4500\\
14  2.7181\\
15	3.0220\\
16  3.3480\\
17  3.6450\\
18	3.9755\\
19  4.3055\\
20  4.6130\\
};
\addlegendentry{$R_{{\rm SD}_{N-1}}$};

\addplot[only marks,mark=square,mark options={},mark size=3pt,draw=green!50!black,fill=green!50!black] plot table[row sep=crcr,]{%
0	6.0576e-05\\
1	0.0021\\
2	0.0231\\
3	0.0968\\
4   0.2604\\
5   0.4801\\
6	0.7212\\
7   0.9036\\
8	0.9818\\ 
9	1.2385\\
10  1.5352\\
11  1.7735\\
12	2.0099\\
13  2.3386\\
14  2.6092\\
15	2.9095\\
16  3.2385\\
17  3.5352\\
18	3.8566\\  
19  4.1933\\
20  4.4971\\
};
\addlegendentry{$R_{{\rm SD}_1}$};

%\addplot[only marks,mark=square,mark options={},mark size=3pt,draw=blue!50!black,fill=blue!50!black] plot table[row sep=crcr,]{%
%0	7.0938e-05\\
%1	0.0025\\
%2	0.0271\\
%3	0.1131\\
%4   0.3049\\
%5   0.5541\\
%6	0.7838\\
%7   0.9302\\
%8	0.9871\\ % updated to here
%9	1.2356\\
%10  1.5320\\
%11  1.7689\\
%12	2.0074\\
%13  2.3386\\
%14  2.6086\\
%15	2.9046\\
%16  3.2356\\
%17  3.5318\\
%18	3.8566\\
%19  4.1933\\
%20  4.4936\\
%};
%\addlegendentry{$R_{{\rm SD}_2}$};

\addplot[only marks,mark=triangle,mark options={},mark size=3pt,draw=orange,fill=orange] plot table[row sep=crcr,]{%
0	6.0576e-05\\
1	0.0021\\
2	0.023\\
3	0.0953\\
4   0.2564\\
5   0.4801\\
6	0.7212\\
7   0.9036\\
8	0.9818\\ 
9	1.2385\\
10  1.5350\\
11  1.7735\\
12	2.0099\\
13  2.3308\\
14  2.6092\\
15	2.9092\\
16  3.2385\\
17  3.5350\\
18	3.8518\\   
19  4.1931\\
20  4.4970\\
};
\addlegendentry{$R_{{\rm SD-BSC}_1}$};

\addplot [color=red,solid]
  table[row sep=crcr]{%
0	0.0413\\
2.5	0.1232\\
5	0.3341\\
7.5	0.7589\\
10	1.3926\\
12.5	2.1477\\
15	2.9526\\
17.5	3.7748\\
20	4.6026\\
};
\addlegendentry{Uniform dist. \cite[(18)]{lmw_bounds}};

\end{axis}
\end{tikzpicture}%

         \caption{Peak constraint only ($E=A$).}
%		\label{capfig}
     \end{subfigure}
     \hspace{.5cm}
     \begin{subfigure}[t]{0.47\textwidth}
         \centering
		\tikzset{every picture/.style={scale=.8}, every node/.style={scale=.8}}
%		\input{rates_vs_SNR_rho_1by3_with_SD_BSC}

% Minimal pgfplots version: 1.3
%  http://www.mathworks.com/matlabcentral/fileexchange/22022-matlab2tikz
\definecolor{mycolor1}{rgb}{0.46600,0.67400,0.18800}%
\begin{tikzpicture}

\begin{axis}[%
width=4.520833in,
height=3.455223in,
xmin=0,
xmax=20,
xlabel={$\frac{{A}}{\sigma}$ [dB]},
xmajorgrids,
ymin=0,
ymax=5,
ylabel={Rate (bits per transmission)},
ymajorgrids,
axis x line*=bottom,
axis y line*=left,
legend style={at={(axis cs:0,5)},anchor=north west,legend cell align=left,align=left,draw=white!15!black}
]

\addplot [color=black,solid,line width=1pt]
  table[row sep=crcr]{%
0	0.144262171974977\\
2.5	0.373895131304245\\
5	0.725101093491492\\
7.5	1.11697769871779\\
10	1.62216615475238\\
12.5	2.22672137916763\\
15	2.90936489704608\\
17.5	3.64660239225238\\
20	4.38108635516173\\
%22.5	5.1722\\
%24  5.5303\\
%25	5.7148\\
%27.5	4.24205056155478\\
%30	3.70044164190238\\
};
\addlegendentry{Capacity};

\addplot[only marks,mark=x,mark options={},mark size=3pt,draw=blue,fill=blue] plot table[row sep=crcr,]{%
     0    0.0001\\
%    0.5000    0.0003\\
    1.0000    0.0018\\
%    1.5000    0.0066\\
    2.0000    0.0193\\
%    2.5000    0.0439\\
3	0.084544787\\
%3.5000    0.1392\\
    4.0000    0.2166\\
%    4.5000    0.3107\\
    5.0000    0.4112\\
%    5.5000    0.5250\\
    6	0.630591264\\
%6.5000    0.7169\\
    7.0000    0.7576\\ 
%    7.5000    0.7983\\
    8.0000    0.8606\\
%    8.5000    0.9286\\
    9.0000    0.9559\\
%    9.5000    0.9774\\
   10.0000    1.1763\\
%   10.5000    1.3152\\
   11.0000    1.5057\\
%   11.5000    1.6004\\      
12	1.7533\\
%12.5000    1.7875\\
   13.0000    1.8487\\
%   13.5000    2.0459\\
   14.0000    2.1704\\
%   14.5000    2.3773\\
15	2.512960555\\
%15.5  2.6370\\
16  2.7281\\
%16.5  2.8793\\
17  3.0626\\
%17.5  3.2310\\
18	3.4099\\
%18.5  3.5477\\
19  3.6332\\
%19.5  3.8164\\
20  3.9665\\
21	4.356233443\\
22 4.6308\\
24	5.326852529\\
27	6.305557138\\
30	7.283524668\\
};
\addlegendentry{$R_{ID}$};

\addplot[only marks,mark=o,mark options={},mark size=3pt,draw=red,fill=red] plot table[row sep=crcr,]{%
         0    0.0001\\
%         .5		.0004\\
         1		.0021\\
%         1.5	.0080\\
         2		.0219\\
%         2.5	.054\\
    3.0000    0.1048\\
%    3.5			.1738\\
    4.0000	  .2717\\
%    4.5			.3672\\
    5			.4794\\
%    5.5			.6060\\
    6.0000    0.7037\\
    7		0.8349\\
    8		0.9310\\
    9.0000    1.2216\\
    10		1.4837\\
    11		1.6988\\
   12.0000    1.9334\\
   13		2.1954\\
   14		2.486\\
   15.0000    2.7607\\
   16		3.0639\\
   17		3.3753\\
   18.0000    3.6715\\
   19		3.9849\\
   20 		4.2841\\
   21.0000    4.6146\\
   24.0000    5.5835\\
   27.0000    6.5755\\
   30.0000    7.4053\\
   };
\addlegendentry{$R_{{\rm SD}_{N-1}}$};

\addplot[only marks,mark=square,mark options={},mark size=3pt,draw=green!50!black,fill=green!50!black] plot table[row sep=crcr,]{%
    0    6.0576e-5\\
    1		0.0018\\
    2	    0.0153\\
    3.0000    0.0672\\ 
    4		0.2232\\ 
    5		0.3931\\
    6.0000    0.6141\\
    7		0.7908\\
    8		0.892\\
    9.0000    1.0798\\
    10			1.3979\\
    11			1.6306\\
   12.0000    1.8441\\
   13			2.1417\\
   14 			2.4198\\
   15.0000    2.6917\\
   16 		3.0016\\
   17		3.2960\\
   18.0000    3.5922\\  % till here
   19		3.9261\\
   20		4.2309\\
   };
\addlegendentry{$R_{{\rm SD}_1}$};

\addplot[only marks,mark=triangle,mark options={},mark size=3pt,draw=orange,fill=orange] plot table[row sep=crcr,]{%
      0    6.0576e-5\\
      1		0.0018\\
      2	    0.0152\\
    3.0000    0.0661\\
    4		0.2175\\
    5		0.3930\\ 
    6.0000    0.6141\\
    7		0.7908\\
    8		0.8919\\
    9.0000    1.079\\
    10			1.396\\
    11			1.6127\\
   12.0000    1.8177\\
   13			2.0998\\
   14 			2.4121\\
   15.0000    2.6904\\
   16 		2.9845\\
   17		3.2793\\
   18.0000    3.5795\\   % till here
   19		3.9083\\
   20		4.2794\\
};
\addlegendentry{$R_{{\rm SD-BSC}_1}$};

\addplot [color=red,solid]
  table[row sep=crcr]{%
  0 0.028886276316062503430306522980557\\
3  0.10871251967552568034850964302105\\
  6  0.36016005610185135704262728423362\\
  9  0.91956324008468704103322246757252\\
 12   1.7467556016003180411989797241752\\
 15    2.693706929352218677719012421044\\
 18   3.6772639562859068470616821992874\\
 20   4.3389917277966257365506238357455\\
};
\addlegendentry{Truncated Exp. dist. \cite[(10)]{lmw_bounds}};

\end{axis}
\end{tikzpicture}%

         \caption{Peak and average constraints, $E=A/3$.}
%		\label{fig_avrgconst}
     \end{subfigure}
    \caption{Achievable rates of the presented coding schemes (Thm. \ref{theorem_independentcodingachievablerate}-\ref{thm_achievablerate1to2}).}
    \label{capfigAllSchemes}
\end{figure}
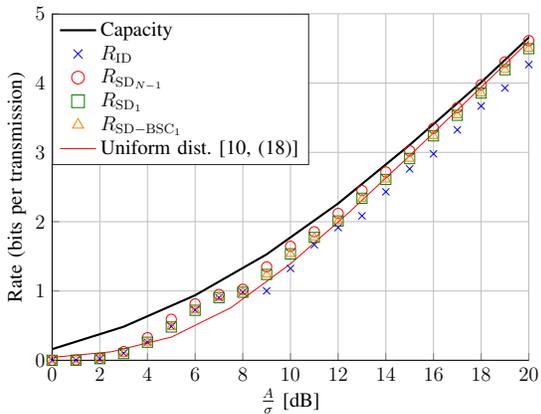
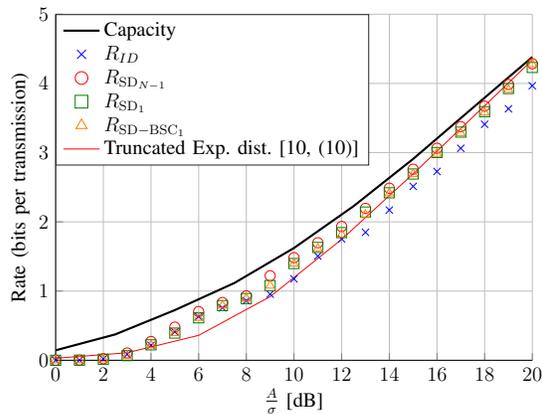

Note that for a given $\mathcal{A}_i$, $R_{{\rm SD}_q}\geq R_{{\rm SD-BSC}_q}$ by the convexity of the mutual information in the transition probability for a given input distribution \cite{cover_elementsIT2}. Numerical evaluations of $R_{{\rm SD}_q}$ are presented in Fig. \ref{capfigAllSchemes}, which shows the importance of exploiting the dependence of $Z_i^b$ on $Z_{i-1}^b,\ldots,Z_0^b$ (cf. Sec. \ref{sec_noise}) at moderate to high SNR, which can enhance the achievable rate. Because of this, the SD$_{N-1}$ scheme achieves rates close to capacity at moderate/high SNR. Note also that the SD$_1$ scheme already improves upon the ID scheme, its rate nearly coincides with that of the SD$_1$ scheme, and it can be implemented using BSC codes.

While the SD scheme closes the gap between the ID scheme and capacity at moderate/high SNR, the gap a low SNR remains. This is because large instances of noise at low SNR `push' information that can help in the decoding process to carry-over bits fed to the next bit-pipes. For instance, if $\gamma A=15$, information can be sent over $4$ bit-pipes ($0,1,2,$ and $3$) due to the peak constraint $\sum_{i\in\mathcal{V}_{\boldsymbol{p}}}2^i\leq\gamma A$, but the received signals on bit-pipes $4,\ldots,N-1$ will contain information from the carry-over from bit-pipes $0,1,2,$ and $3$. To exploit this, a scheme which decodes $m_i$ from $Y_i^c,\ldots,Y_{i+q-1}^c$ to recover information from the carry-over bits is proposed. 

\subsection{Carry-Over-Assisted Decoding (CD)}
\label{sec_bigob2t}
In this scheme, which will be called the CD$_q$ scheme, the  $m_i$ is decoded using the output of bit-pipe $i$ and the output of the next $q$ bit-pipes which depend on carry-over information of interest. Its achievable rate is given next, and a simplification which enables using BCC while exploiting the carry-over from the next $q$ bit-pipes is proposed afterwards.

\subsubsection{The CD$_q$ Scheme}
The receiver starts decoding from the $q+1$ lowermost bit-pipes and proceeds to higher bit-pipes. For each decoding operation, the receiver decodes $m_i$ from the output $(\bar{Y}_{i}^c(t),Y_{i+1}^c(t),\ldots,Y_{i+q}^c(t))$ for $t=1,2,\ldots,n$ (where $Y_i^c=0$ for $i>N-1$ and $\bar{Y}_i^c(t)$ is defined as in \eqref{Yibar}). The transition probability of such a channel can be calculated using the binary noise probability distribution (Sec. \ref{sec_noise}) and our knowledge of $p_i$ (a design parameter). This scheme is a simplified version of MLC/MSD scheme with a quantized output, where $q+1$ bit-pipes are considered for decoding each stream of bits instead of all $N$ bit-pipes. Thus this scheme is less complex but is expected to have lower performance. Next, the achievable rate of this scheme is expressed.

\begin{theorem}[Rate of CD$_q$]
\label{thm_achievablerate1to2}
For given $q\in\{1,\ldots,N-1\}$, $\beta$, and $\gamma$, the CD$_q$ scheme achieves
 $R_{\text{CD}_q}   = (1-\bar{\epsilon}) \max_{p_i} \sum_{i=0}^{N-1} I(X_i^b;\bar{Y}_i^c,Y_{i+1}^c,\ldots,Y_{i+q}^c|\bar{Y}_i^c\neq\xi)$, where $Y_N^c=Y_{N+1}^c=\ldots=0$, and where $p_i$ satisfies $\sum_{i\in \mathcal{V}_{\boldsymbol{p}}}2^i\leq \gamma A$ and $\sum_{i\in \mathcal{V}_{\boldsymbol{p}}}p_i2^i\leq \gamma E$. 
\end{theorem}
\begin{IEEEproof}
Bit-pipe $i$, for $i=0,\ldots,N-1$ is a channel with binary input $X_i^b$ and $(q+1)$-ary output $\bar{Y}_i^c,Y_{i+1}^c,\ldots,Y_{i+q}^c$, concatenated with a BEC$(\epsilon)$. Using Lemma \ref{lem_erasure}, the rate $R_i=(1-\epsilon)I(X_i^b;\bar{Y}_i^c, Y_{i+1}^c,\ldots Y_{i+q-1}^c|\bar{Y}_i^c\neq \xi)$ is achievable over this bit-pipe. Noting that $\epsilon \leq \bar{\epsilon}$ \eqref{eq_perasure}, and summing over all $i$ and maximizing with respect to $p_i$, the total rate $R_{\text{CD}_q}$ is obtained.
\end{IEEEproof}

In general, using the additional outputs $Y_{i+1}^c,\ldots,Y_{i+q}^c$ in the DC$_q$ scheme can improve the achievable rate compared to the ID scheme since $I(X_i^b;\bar{Y}_i^c,Y_{i+1}^c,\ldots Y_{i+q}^c|\bar{Y}_i^c\neq\xi)$ can be written as $I(X_i^b;\bar{Y}_i^c|\bar{Y}_i^c\neq\xi)+(1-\bar{\epsilon})I(X_i^b;Y_{i+1}^c,\ldots,Y_{i+q}^c|\bar{Y}_i^c,\bar{Y}_i^c\neq \xi)$ by the chain rule, which is larger than $I(X_i^b;\bar{Y}_i^c|\bar{Y}_i^c\neq\xi)$ since mutual information is nonnegative. Similar to before, a simplification that enables using BCC is proposed next.

\subsubsection{The CD-BAC$_q$ Scheme}    
Similar to the SD-BSC scheme, the outputs of bit-pipes $i+1,\ldots,i+q$ can be used to flip the output of bit-pipe $i$ before using it for decoding. This converts the channel governing the CD scheme into a binary channel where BCC can be used. Since the resulting binary channel is asymmetric as shall be shown next, this scheme will be denoted the CD-BAC scheme (binary asymmetric channel). Let us start by describing the case $q=1$.

Consider bit-pipe $i$, and suppose that $\mathcal{P}\{(X_{i+1}^b+Z_{i+1}^b)\bmod 2=1|Y_{i+1}^c\neq \xi\}=\theta_{i}$ and $\bar{\theta}_i=1-\theta_i$. Note that $X_{i+1}^b$ contains information about the carry-over produced by bit-pipe $i$. The quantity $\mathcal{P}\{Y_{i+1}^b=1|Y_{i}^b=0\}$ will be of interest, since this event will be used to decide whether to flip $Y_i^b$ when $Y_i^b=0$. Note that the cases leading to $Y_i^b=0$ can be classified into carry-over cases where $(X_i^b,Z_i^b)=(1,1)$, and non-carry-over cases where $(X_i^b,Z_i^b)=(0,0)$.

Note that $\mathcal{P}\{Y_{i+1}^b=1|Y_{i}^b=0\}=\theta_i(1-p_i)(1-\alpha_i)+\bar{\theta}_ip_i\alpha_i$, where $p_i=\mathcal{P}\{X_i^b=1\}$ and $\alpha_i=\mathcal{P}\{Z_i^b=1\}$ are both known at the transmitter and receiver. Thus, a fraction $\bar{\theta}_i$ of the carry-over cases results in $Y_{i+1}^b=1$, and a fraction $\theta_i$ of the non-carry-over cases results in $Y_{i+1}^b=1$. If $\theta_i<0.5$, a majority of the carry-over cases results in $Y_{i+1}^b=1$ and a minority of the non-carry-over cases results in $Y_{i+1}^b=1$. Thus, if the receiver flips $Y_i^b$ when $(Y_{i+1}^b,Y_i^b)=(1,0)$, this corrects a fraction $\bar{\theta}_i$ (a majority) of the carry-over cases, and flips (into an error) a fraction $\theta_i$ (a minority) of the non-carry-over cases. This process of correcting some output bits introduces errors to others as quantified next. Let us denote the flipped $Y_i^b$ by $\check{Y}_i^b$. Then,
\begin{align}
\label{Ybreve01}
\mathcal{P}\{\check{Y}_i^b=0|X_i^b=1\}&=\alpha_i-\alpha_i\bar{\theta}_i,\\
\label{Ybreve10}
\mathcal{P}\{\check{Y}_i^b=1|X_i^b=0\}&=\alpha_i+(1-\alpha_i)\theta_i,
\end{align}
where the subtracted term accounts for carry-over cases that have been corrected, and the added term accounts for non-carry-over cases that have been erroneously flipped. The resulting channel is asymmetric, but can have higher capacity than the original channel BSC$(\alpha_i)$ depending on $\theta_i$. One extreme example is when $\theta_i=0$, where the new channel is a Z-channel with crossover probability $\alpha_i$ which has higher capacity than BSC$(\alpha_i)$.

Let us now consider first what happens when $\theta_i>0.5$. In this case, the description above will correct a minority of carry-over cases, and flip (into an error) a majority of non-carry-over cases. However, the receiver can flip $Y_i^b=0$ to $1$ if $Y_{i+1}^b=0$ instead. Since $\mathcal{P}\{(X_{i+1}^b+Z+{i+1}^b)\bmod 2=0\}<0.5$ in this case, the scheme works as above, and the crossover probabilities of the resulting BAC are as in \eqref{Ybreve01} and \eqref{Ybreve10} with $\theta_i=\mathcal{P}\{(X_{i+1}^b+Z_{i+1}^b)\bmod 2=0\}$.

This scheme can be generalized for $q\geq1$ as follows. Let $\boldsymbol{T}_i^b=(Y_{i+1}^b,\ldots,Y_{i+q}^b)$ and let
\begin{align}
\label{theta_i}
\theta_i&=\mathcal{P}\{\boldsymbol{T}_i^b\in\mathcal{T}_i|(X_i^b,Z_i^b)=(0,0)\},\\
\label{theta_i_bar}
\bar{\theta}_i&=\mathcal{P}\{\boldsymbol{T}_i^b\in\mathcal{T}_i|(X_i^b,Z_i^b)=(1,1)\},
\end{align}
where $\mathcal{T}_i\subseteq\{0,1\}^q$, and where $Y_i^b=0$ for $i>N-1$. Then, the receiver flips $Y_i^b=0$ to $1$ when $\boldsymbol{T}_i^b\in\mathcal{T}_i$, and the cross-over probability of the resulting BAC can be described as in \eqref{Ybreve01} and \eqref{Ybreve10} with $\theta_i$ and $\bar{\theta}_i$ as defined in \eqref{theta_i} and \eqref{theta_i_bar}. The achievable rate of this scheme follows next.

\begin{theorem}[Rate of CD-BAC$_q$]
\label{thm_CD_BAC}
For given $q\in\{1,\ldots,N-1\}$, $\beta$, $\gamma$, and sets $\mathcal{T}_i\subseteq\{0,1\}^q$,  $i=0,\ldots,N-1$, the CD-BAC$_q$ scheme achieves the rate $R_{{\rm CD-BAC}_q} = (1-\bar{\epsilon}) \max_{p_i} \sum_{i=0}^{N-1} \Phi_i$ where $\Phi_i=\left(\mathsf{h}((1-p_i)\tilde{\alpha}_{i0}+p_i(1-\tilde{\alpha}_{i1}))-(1-p_i)\mathsf{h}(\tilde{\alpha}_{i0})-p_i\mathsf{h}(\tilde{\alpha}_{i0})\right)$, $\tilde{\alpha}_{i0}=\alpha_i+(1-\alpha_i)\theta_i$, $\tilde{\alpha}_{i1}=\alpha_i-\alpha_i\bar{\theta}_i$, $\theta_i$ and $\bar{\theta}_i$ are defined in \eqref{theta_i} and \eqref{theta_i_bar}, and $p_i$ satisfies $\sum_{i\in \mathcal{V}_{\boldsymbol{p}}}2^i\leq \gamma A$ and $\sum_{i\in \mathcal{V}_{\boldsymbol{p}}}p_i2^i\leq \gamma E$. 
\end{theorem}
\begin{IEEEproof}
After flipping $Y_i^b$ into $\check{Y}_i^b$ whose probability is described in \eqref{Ybreve01} and \eqref{Ybreve10}, the resulting BAC has capacity given by $I(X_i^b;\check{Y}_i^b)=\mathsf{h}((1-p_i)\tilde{\alpha}_{i0}+p_i\tilde{\alpha}_{i1})-(1-p_i)\mathsf{h}(\tilde{\alpha}_{i0})-p_i\mathsf{h}(\tilde{\alpha}_{i0})$. Observing that the overall channel is concatenated with a BEC$(\epsilon)$ with $\epsilon\leq\bar{\epsilon}$ \eqref{eq_perasure}, summing over all $i$, and maximizing with respect to $p_i$ leads to $R_{{\rm CD-BAC}_q}$.
\end{IEEEproof}

Note that $\tilde{\alpha}_{i0}+\tilde{\alpha}_{i1}=2\alpha_i+(1-\alpha_i)\theta_i-\alpha_i\bar{\theta}_i$. Thus, a good selection of $\mathcal{T}_i$ is one which results in $(1-\alpha_i)\theta_i-\alpha_i\bar{\theta}_i<0$ when $\alpha_i<0.5$ and $(1-\alpha_i)\theta_i-\alpha_i\bar{\theta}_i>0$ otherwise, so that the new channel introduces less errors than BSC$(\alpha_i)$ and BSC$(1-\alpha_i)$, respectively, when $p_i=0.5$. This scheme can be implemented in practice using BCC for asymmetric channels, such as \cite{NagarjunaSiegel,HondaYamamoto}.

\begin{figure}
    \centering
\begin{subfigure}[t]{0.47\textwidth}
         \centering
         \tikzset{every picture/.style={scale=.8}, every node/.style={scale=.8}}
%         \input{rates_vs_SNR_peak_only}%

% Minimal pgfplots version: 1.3
%  http://www.mathworks.com/matlabcentral/fileexchange/22022-matlab2tikz
\definecolor{mycolor1}{rgb}{0.38824,0.15686,0.15686}%
\begin{tikzpicture}

\begin{axis}[%
width=4.520833in,
height=3.458695in,
xmin=0,
xmax=10,
xlabel={$\frac{{A}}{\sigma}$ [dB]},
xmajorgrids,
ymin=0,
ymax=2,
ylabel={Rate (bits per transmission)},
ymajorgrids,
axis x line*=bottom,
axis y line*=left,
legend style={at={(axis cs: 0,2)},anchor=north west,legend cell align=left,align=left,draw=white!15!black}
]

\addplot [color=black,solid,line width=1pt]
  table[row sep=crcr]{%
0	0.160700000121326\\
3	0.484399999620577\\
6	0.938500000064862\\
9	1.52889999948447\\
12	2.26260000038404\\
};
\addlegendentry{Capacity};

\addplot[only marks,mark=x,mark options={},mark size=3pt,draw=blue,fill=blue] plot table[row sep=crcr,]{%
0	6.04709954474166e-05\\
1   0.0021\\
2   0.0225\\
3	0.101600000656657\\
4   .2597\\
5   .4879\\
6	0.727200000428726\\
7   .903\\
8	.9805\\
9	1.00219999948542\\
10  1.3248\\
};
\addlegendentry{$R_{\rm ID}$};

\addplot[only marks,mark=o,mark options={},mark size=3pt,draw=red,fill=red] plot table[row sep=crcr,]{%
0	7.4655e-05\\
1	0.0027\\
2	0.0280\\
3	0.1174\\
4   0.3261\\
5   0.5916\\
6	0.8135\\
7   0.9483\\
8	1.0221\\
9	1.3480\\
10  1.6450\\
};
\addlegendentry{$R_{{\rm SD}_{N-1}}$};

%\addplot[only marks,mark=square,mark options={},mark size=3pt,draw=green!50!black,fill=green!50!black] plot table[row sep=crcr,]{%
%0	7.4655e-05\\
%1	0.0022\\
%2	0.0227\\
%3	0.0992\\
%4   0.2638\\
%5   0.4880\\
%6	0.7254\\
%7   0.9002\\
%8	0.9810\\
%9	1.2356\\
%10  1.5320\\
%};
%\addlegendentry{$R_{{\rm SD}_1}$};

\addplot[only marks,mark=diamond,mark options={},mark size=3pt,draw=green!50!black,fill=green!50!black] plot table[row sep=crcr,]{%
      0    0.0531\\
      1		0.1401\\
      2	    0.2628\\
    3.0000    0.3944\\
    4		0.5327\\
    5		0.7111\\
    6.0000    0.8637\\
    7		0.9339\\
    8		1.0561\\
    9.0000    1.2365\\
    10			1.4249\\
};
\addlegendentry{$R_{{\rm CD}_1}$};

%\addplot[only marks,mark=square,mark options={},mark size=3pt,draw=green!50!black,fill=green!50!black] plot table[row sep=crcr,]{%
%      0    0.1777\\ %OK
%      1		0.2423\\ %OK
%      2	    0.3368\\ %OK
%    3.0000    0.4362\\ %OK
%    4		0.5560\\ %OK
%    5		0.7107\\ %OK
%    6.0000    0.8637\\  % OK
%    7		0.9352\\ % OK
%    8		1.0561\\
%    9.0000    1.2365\\
%    10			1.4249\\
%};
%\addlegendentry{$R_{{\rm CD}_2}$};

\addplot[only marks,mark=*,mark options={},mark size=2pt,draw=orange,fill=orange] plot table[row sep=crcr,]{%
      0    0.0337\\
      1		0.0865\\
      2	    0.1754\\
    3.0000    0.2948\\
    4		0.4726\\
    5		0.6324\\
    6.0000    0.7838\\ 
    7		0.9290\\ 
    8		0.9841\\
    9.0000    1.1178\\
    10			1.4241\\% updated to here
};
\addlegendentry{$R_{{\rm CD-BSC}_1}$};

\addplot [color=red,solid]
  table[row sep=crcr]{%
0	0.0413\\
2.5	0.1232\\
5	0.3341\\
7.5	0.7589\\
10	1.3926\\
%12.5	2.1477\\
%15	2.9526\\
%17.5	3.7748\\
%20	4.6026\\
};
\addlegendentry{Uniform dist. \cite[(18)]{lmw_bounds}};

\end{axis}
\end{tikzpicture}%

         \caption{Peak constraint only ($E=A$).}
%		\label{capfig}
     \end{subfigure}
     \hspace{.5cm}
     \begin{subfigure}[t]{0.47\textwidth}
         \centering
		\tikzset{every picture/.style={scale=.8}, every node/.style={scale=.8}}
%		\input{rates_vs_SNR_rho_1by3}

% Minimal pgfplots version: 1.3
%  http://www.mathworks.com/matlabcentral/fileexchange/22022-matlab2tikz
\definecolor{mycolor1}{rgb}{0.46600,0.67400,0.18800}%
\begin{tikzpicture}

\begin{axis}[%
width=4.520833in,
height=3.455223in,
xmin=0,
xmax=10,
xlabel={$\frac{{A}}{\sigma}$ [dB]},
xmajorgrids,
ymin=0,
ymax=2,
ylabel={Rate (bits per transmission)},
ymajorgrids,
axis x line*=bottom,
axis y line*=left,
legend style={at={(axis cs:0,2)},anchor=north west,legend cell align=left,align=left,draw=white!15!black}
]
\addplot [color=black,solid,line width=1pt]
  table[row sep=crcr]{%
0	0.144262171974977\\
2.5	0.373895131304245\\
5	0.725101093491492\\
7.5	1.11697769871779\\
10	1.62216615475238\\
12.5	2.22672137916763\\
};
\addlegendentry{Capacity};

\addplot[only marks,mark=x,mark options={},mark size=3pt,draw=blue,fill=blue] plot table[row sep=crcr,]{%
     0    0.0001\\
    1.0000    0.0018\\
    2.0000    0.0193\\
3	0.084544787\\
    4.0000    0.2166\\
    5.0000    0.4112\\
    6	0.630591264\\
    7.0000    0.7576\\ 
    8.0000    0.8606\\
    9.0000    0.9559\\
   10.0000    1.1763\\
};
\addlegendentry{$R_{\rm ID}$};

\addplot[only marks,mark=o,mark options={},mark size=3pt,draw=red,fill=red] plot table[row sep=crcr,]{%
         0    0.0001\\
         1		.0021\\
         2		.0219\\
    3.0000    0.1048\\
    4.0000	  .2717\\
    5			.4794\\
    6.0000    0.7037\\
    7		0.8349\\
    8		0.9310\\
    9.0000    1.2216\\
    10		1.4837\\
   };
\addlegendentry{$R_{{\rm SD}_{N-1}}$};

%\addplot[only marks,mark=square,mark options={},mark size=3pt,draw=green!50!black,fill=green!50!black] plot table[row sep=crcr,]{%
%    0    0.000050809\\
%    1		0.0019\\
%    2	    0.0195\\
%    3.0000    0.0863\\
%    4		0.2261\\
%    5		0.4158\\
%    6.0000    0.6313\\
%    7		0.7920\\
%    8		0.8990\\
%    9.0000    1.1135\\
%    10			1.3979\\
%   };
%\addlegendentry{$R_{{\rm SD}_1}$};

\addplot[only marks,mark=diamond,mark options={},mark size=3pt,draw=green!50!black,fill=green!50!black] plot table[row sep=crcr,]{%
      0    0.0531\\
      1		0.1408\\
      2	    0.2641\\
    3.0000    0.3944\\
    4		0.5270\\
    5		0.6880\\
    6.0000    0.8411\\
    7		0.9461\\
    8		1.0805\\
    9.0000    1.2403\\
    10			1.4067\\
};
\addlegendentry{$R_{{\rm CD}_1}$};

\addplot[only marks,mark=*,mark options={},mark size=2pt,draw=orange,fill=orange] plot table[row sep=crcr,]{%
      0    0.0293\\
      1		0.0794\\
      2	    0.1558\\
    3.0000    0.2633\\ 
    4		0.3739\\
    5		0.6078\\ 
    6.0000    0.7265\\ 
    7		0.7883\\ 
    8		0.8526\\
    9.0000    0.9717\\
    10			1.1546\\ % updated
};
\addlegendentry{$R_{{\rm CD-BSC}_1}$};

\addplot [color=red,solid]
  table[row sep=crcr]{%
  0 0.028886276316062503430306522980557\\
3  0.10871251967552568034850964302105\\
  6  0.36016005610185135704262728423362\\
  9  0.91956324008468704103322246757252\\
 12   1.7467556016003180411989797241752\\
};
\addlegendentry{Truncated Exp. dist. \cite[(10)]{lmw_bounds}};

\end{axis}
\end{tikzpicture}%

         \caption{Peak and average constraints, $E=A/3$.}
%		\label{fig_avrgconst}
     \end{subfigure}
    \caption{Achievable rates of the presented coding schemes (Thm. \ref{theorem_independentcodingachievablerate}-\ref{thm_achievablerate1to2}).}
    \label{CD}
\end{figure}
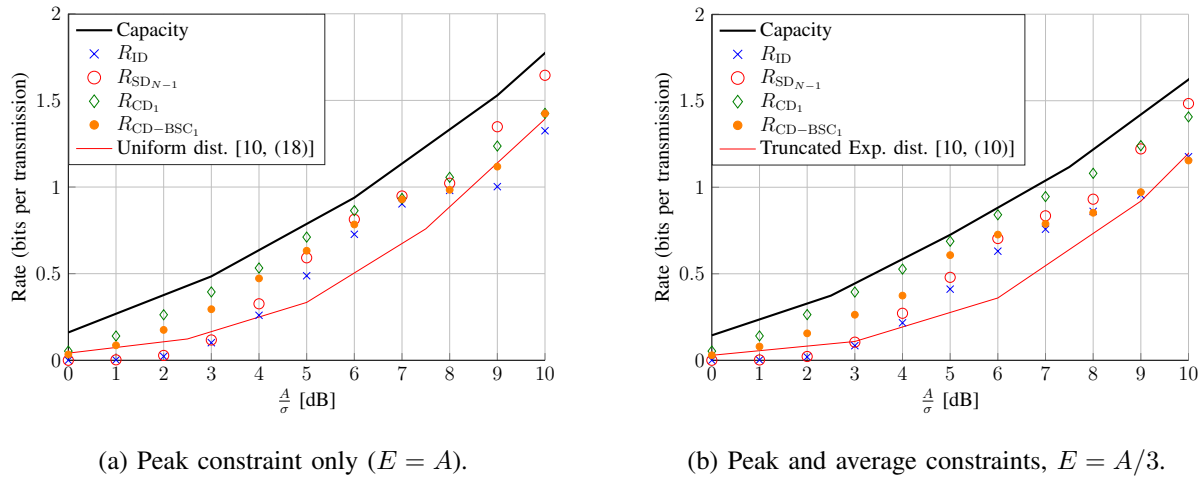

Fig. \ref{CD} shows numerical evaluations of the achievable rates of the CD$_1$ and CD-BAC$_1$ schemes, focusing on the low/medium SNR range. Note that at low SNR, exploiting information on higher bit-pipes using the CD$_1$ scheme is important, because the noise in this regime leads to sufficient dependence between $X_i^b$ and the output of higher bit-pipes (through carry-over bits). This leads to higher rates for the CD$_1$ and CD-BAC$_1$ schemes than the ID and SD schemes. Higher rates are expected with $q>1$.

\section{Discussion and Implementation}
\label{sec_resultsanddiscussion}
In this section, the advantages and disadvantages of the presented VBC decomposition and decoding schemes are discussed, their implementation using polar codes is evaluated.

\subsection{Discussion of Results}
The presented VBC decomposition results in a channel with input $(X_0^b,\ldots,X_{N-1}^b)$ and output $(Y_0^c,\ldots,Y_{N-1}^c)$. This provides an alternative perspective into IM/DD channels viewed as a set of bit-pipes. The decomposition does not degrade the channel, since the capacity of the resulting VBC approaches that of the original IM/DD channel when $\beta$ and $\gamma$ are large enough (Thm. \ref{CapacityPreserving}). The achievable rates shown in Fig. \ref{capfigfinal}-\ref{CD} deviate from capacity due to the following reasons.

At the transmitter, to achieve capacity, the distribution of $X$ should coincide with the optimal input distribution. The independent encoding scheme cannot generate the optimal input distribution in general. This is because the generated $X$ has an alphabet size which is a power of $2$, and the mass points are equally spaced between $0$ and $A$. Moreover, even if the optimal input alphabet satisfies this, the encoder may not be able to generate the optimal distribution. In fact, the generated distribution obtained by the considered encoder is constrained to the form $P_{0}(x)\otimes P_1(x)\otimes \cdots\otimes P_{N-1}(x)$, where $\otimes$ is the convolution operator, $P_i(x)=(1-p_i)\delta(x)+p_i\delta(x-2^i/\gamma)$, and $\delta(x)$ is the Dirac function. This limits the shaping of the input distribution. For example, at $A=6.3$ ($A/\sigma\approx8$ dB) with no average constraint, the optimal input distribution is given by $(0.34,0.16,0.16,0.34)$ with support $(0,2.58,3.78,6.3)$, which cannot be generated from the convolution above. In fact, calculating the binary representation of the optimal $X$ as $(X_0^b,\ldots,X_{N-2}^b,X_{N-1}^b)$ shows that $P\{X_{N-2}^b=1|X_{N-1}^b=0\}=0.16$ and $P\{X_{N-2}^b=1|X_{N-1}^b=1\}=0.34$. %the second most significant bit is $1$ with probability $0.16$ when the most significant bit is $0$, and with probability $0.34$ when the most significant bit is $1$.
 This dependence cannot be generated by independent encoding, it requires encoding the bits $(X_0^b,\ldots,X_{N-1}^b)$ jointly. Nevertheless, this encoding still achieves rates close to capacity when combined with MSD as seen from Fig.~\ref{capfigfinal}.

At the receiver, the ID scheme is simple, can be implemented using BCC (such as LDPC, polar, or turbo codes), but does not exploit inherent dependence between the outputs of bit-pipes which sacrifices capacity. Both the SD and CD schemes exploit such dependence, which recovers some capacity loss, and both admit a simple implementation using BCC (the SD-BSC and the CD-BAC schemes). The SD and CD schemes can be implemented using polar codes for binary input channels (for these schemes, the output is not binary but a binary vector). Since these schemes result in asymmetric channels in general, shaping may be needed and can be implemented as in \cite{HondaYamamoto}. The SD-BSC and the CD-BAC can be implemented using polar codes for BIBO channels, where shaping is needed for the CD-BAC scheme since the channel for each bit-pipe can be asymmetric in general. However, all these schemes achieve rates lower than capacity since they do not exploit all dependencies between bit-pipes as explained earlier.

In general, the proposed schemes can be seen as degraded versions of MSD, which exploits all dependencies at the output. Thus, MLC/MSD is expected to perform better than the schemes above in general. Exploiting all dependencies at the receiver side can be obtained using our proposed schemes by combining the SD and CD schemes, but this path has not been attempted in this work. Moreover, performance improvements can be achieved by considering shaping, custom designed mappers, or distribution matchers as in \cite{prob_4pam,tw_A,tw_B,tw_C,fh_signalling}.

Next, numerical simulations of an implementation using polar coding are presented.

\subsection{Polar Coding Over the VBC}
\label{sec_polarindependent}
The purpose of this section is to validate the theoretical results in Sec. \ref{Sec:BCC} by employing BCC over the VBC. Polar codes are selected as the BCC since they achieve the capacity of binary input discrete memoryless channels (Bi-DMC) \cite{arikan_channelpolarization,
erdal_ontheoriginofpolarcodes}. But other BCC (such as LDPC or turbo codes) can be used as well. This section proposes a practical implementation of the ID and SD-BSC schemes using polar codes. The resulting achievable rates, bit-error rate (BER), and frame error rate (FER) are investigated. Specifically, the goal is to use BSC polar encoders and decoders for the channel in Fig. \ref{fig_imddchannelbp} and simulate the performance, in order to validate the theoretical results presented in Fig. \ref{capfigAllSchemes}. This is done as follows. 

To implement the ID scheme, the capacity of an active bit-pipe $i\in\mathcal{V}_{\boldsymbol{p}}$ modelled as a BSC$(\alpha_i)$ is first evaluated. Then the number of information bits $k_i$ is chosen as the code-length $n$ multiplied by this capacity minus a margin $\zeta_i$, i.e., $k_i=n(1-\mathsf{h}(\alpha_i)-\zeta_i)$. This back-off term $\zeta_i$ is needed since we will use finite-length codes, and the maximum achievable rate over a channel in the finite block-length regime is less than the Shannon capacity \cite{Polyanskiy}. Then, a polar code with rate $k_i/n$ and length $n$ for a BSC$(\alpha_i)$ is used to generate the codeword $(X_i^b(1),\ldots,X_i^b(n))$ from a message $m_i$ with $k_i$ bits. Then, a polar decoder is used to decode $\hat{m}_i$ from $(Y_i^c(1),\ldots,Y_i^c(n))$. Recall that the scheme reconstructs $W_{i+1}^b(t)$ which is a function of $X_i^b(t)$. Decoding errors will lead to carry-over reconstruction errors which are considered in the simulation. Finally, the decoding outcome is used to evaluate the bit-error rate (BER) and the frame error rate (FER). A similar procedure is used for the SD-BSC scheme.

In the simulation, $\bar{T}$ frames are transmitted, each of length $n$ symbols over each of the $N$ bit-pipes. Then, the BER over bit-pipe $i$ is evaluated as the ratio of the number of information bit errors on bit-pipe $i$ to the total number of sent information bits $k_i\bar{T}$, i.e., $\text{BER}_i=\frac{1}{k_i\bar{T}}\sum_{j=1}^{\bar{T}} \sum_{k=1}^{k_i} \mathsf{I}_ {U_{i,j}^b(k)\neq \hat{U}_{i,j}^b(k)}$,
where $U_{i,j}^b(k)$ and $\hat{U}_{i,j}^b(k)$ represent the $k^\textnormal{th}$ bit in the message $m_i(j)$ and $\hat{m}_i(j)$ sent and decoded in frame $j$, respectively. The overall BER over all bit-pipes is obtained using $\text{BER}=\frac{1}{\sum_{i\in\mathcal{V}_{\boldsymbol{p}}}k_i\bar{T}}\sum_{i\in\mathcal{V}_{\boldsymbol{p}}}\sum_{j=1}^{\bar{T}} \sum_{k=1}^{k_i} \mathsf{I}_ {U_{i,j}^b(k)\neq \hat{U}_{i,j}^b(k)}$.

To calculate the FER over bit-pipe $i$, the ratio of the count of erroneous frames to the total number of sent frames is calculated, i.e., $\text{FER}_i= \frac{1}{\bar{T}} \sum_{j=1}^{\bar{T}} \left(1-\prod_{k=1}^{k_i} \mathsf{I}_{U_{i,j}^b(k)= \hat{U}_{i,j}^b(k)}\right)$. This frame error rate is important if independent information streams are transmitted in parallel across the bit-pipes. On the other hand, if a message $m$ is split into messages $m_i$ and sent across the bit-pipe, then a frame error corresponds to an error in $m$, which takes place whenever one or more bit-pipes have an erroneous decoding. Thus, the overall FER in this case is $\text{FER}= \frac{1}{\bar{T}} \sum_{j=1}^{\bar{T}} \left(1-\prod_{i\in\mathcal{V}_{\boldsymbol{p}}}\prod_{k=1}^{k_i} \mathsf{I}_{U_{i,j}^b(k)= \hat{U}_{i,j}^b(k)}\right)$. These quantities are evaluated numerically next.

Tab. \ref{tab_examplepolarcodee} shows an example for the ID scheme over an IM/DD channel with $A=25$, $\gamma=4.4801$, and $\beta=5$, under $\bar{T}=1000$ frames, where $\gamma$ and $\beta$ were optimized using grid search over $\beta\in[3,5]$ and $\gamma\in[3,6]$. For each bit-pipe $i$, the table shows its cross-over probability $\alpha_i$, capacity $1-\mathsf{h}(\alpha_i)$, rate of the used polar code $k_i/n$, and the BER and FER. Active bit-pipes are chosen such that the peak constraint is satisfied. Encoding and decoding are done using the polar code packages in \cite{vangala2016practical,PolarCodeGitHub}. The resulting BER and FER are small, while the achievable rate is within $\approx 0.3$ bits of the theoretical achievable rate of the ID scheme.\footnote{This gap is mainly due to the use of finite codes (cf. \cite[Fig. 1]{Polyanskiy}. For instance, at $n=64$, the gap between the bit-pipe's (BSC) capacity and the maximum rate in the finite blocklength regime is $\sim 10^{-1}$ when the FER is $\sim10^{-2}$.} One can increase the achievable rate by increasing $k_i$ (for fixed $n$) if higher BER/FER is allowed, or by using other BCC such as LDPC or Reed-Muller codes which can perform better with a small~$n$.

Tab. \ref{tab_examplepolarcodee_SD} shows the results for the SD-BSC$_1$ scheme. Note how $\breve{\alpha}_i$ is lower than $\alpha_i$, resulting in achievable rates that are higher than the ID scheme.

\begin{table}[t]
\centering
\begin{tabular}{c||c|c"c|c|c"c|c|c}
\hline
\multirow{2}{*}{$i$} & \multirow{2}{*}{$\min\{\alpha_i,1-\alpha_i\}$} & \multirow{2}{*}{$1-\mathsf{h}(\alpha_i)$} & \multicolumn{3}{c"}{$n=2^{20}$} & \multicolumn{3}{c}{$n=2^6$} \\\cline{4-9}
& & & $k_i/n$ &  $\text{BER}_i$ & $\text{FER}_i$ & $k_i/n$ &  $\text{BER}_i$ & $\text{FER}_i$ \\ \hline\hline
$0-3$ & $>0.42$ & $<0.018$ & $0$ & $0$ & $0$  & $0$ & $0$ & $0$ \\ \hline
%$1$ & $0.5$ & $0$ & $0$ & $0$ & $0$  & $0$ & $0$ & $0$\\ \hline
%$2$ & $0.498$ & $4\mathrm{e}{-6}$& $0$ & $0$ & $0$ & $0$ & $0$ & $0$\\ \hline
%$3$ & $0.420$& $0.018$ & $0$ & $0$ & $0$ & $0$ & $0$ & $0$\\ \hline
$4$ & $0.092$ & $0.554$ & $0.399$ & $0$ & $0$ & $0.359$ & $0.025$ & $0.077$\\ \hline
$5$ & $0.016$ & $0.881$ & $0.722$ & $0$ & $0$ & $0.734$ & $0.026$ & $0.090$\\ \hline
$6$ & $3\mathrm{e}{-7}$ & $0.999$ & $0.999$ & $8\mathrm{e}{-4}$ & $4\mathrm{e}{-2}$ & $0.984$ & $0.018$ & $0.093$\\ \hline
$7$ & $3\mathrm{e}{-7}$ & $0.999$ & $0$ & $0$ & $0$ & $0$ & $0$ & $0$\\ \hline\hline
\multicolumn{2}{c|}{Overall (active)}& $2.436$ & $2.122$ & $8\mathrm{e}{-4}$ & $4\mathrm{e}{-2}$& $2.015$ & $0.022$ & $0.095$\\
\end{tabular}
\caption{BER and FER for ID with polar coding: $A=25$, $\gamma=4.4801$, $\beta=5$. The rate ($k_i/n$) is tuned to be as close as possible to the capacity of the bit-pipe while achieving small FER. Active bit-pipes are chosen to achieve the highest rate while satisfying the peak constraint. Bit-pipe $7$ is not active ($p_7=0$) because $2^7=128>\gamma A=112.002$ which violates the peak constraint, while bit-pipes $4$, $5$, and $6$ are active because $2^4+2^5+2^6=112<\gamma A$.}
\label{tab_examplepolarcodee}
\end{table}

\begin{table}[t]
\centering
\begin{tabular}{c||c|c|c"c|c|c"c|c|c}
\hline
\multirow{2}{*}{$i$} & \multirow{2}{*}{$\min\{\alpha_i,1-\alpha_i\}$} & \multirow{2}{*}{$\breve{\alpha}_i$ (Thm. \ref{theorem_SD_BSC})} & \multirow{2}{*}{$1-\mathsf{h}(\breve{\alpha}_i)$} & \multicolumn{3}{c"}{$n=2^{20}$} & \multicolumn{3}{c}{$n=2^6$} \\\cline{5-10}
& & & & $k_i/n$ &  $\text{BER}_i$ & $\text{FER}_i$ & $k_i/n$ &  $\text{BER}_i$ & $\text{FER}_i$ \\ \hline\hline
$0-3$ & $>0.483$ & $>0.365$ & $<0.053$ & $0$ & $0$ & $0$  & $0$ & $0$ & $0$ \\ \hline
%$1$ & $0.5$ & $0.5$ & $0$ & $0$ & $0$ & $0$  & $0$ & $0$ & $0$\\ \hline
%$2$ & $0.499$ & $0.498$ & $4\mathrm{e}{-6}$ & $0$ & $0$ & $0$ & $0$ & $0$ & $0$\\ \hline
%$3$ & $0.499$& $0.365$ & $-$ & $0$ & $0$ & $0$ & $0$ & $0$ & $0$\\ \hline
$4$ & $0.4717$ & $0.075$ & $0.615$ & $0.45$ & $5\mathrm{e}{-6}$ & $0.05$ & $0.359$ & $0.009$ & $0.036$ \\ \hline
$5$ & $4\mathrm{e}{-4}$ & $4\mathrm{e}{-4}$ & $0.995$ & $0.916$ & $5\mathrm{e}{-5}$ & $0.04$  & $0.843$ & $0.011$ & $0.035$ \\ \hline
$6$ & $2\mathrm{e}{-4}$ & $1\mathrm{e}{-12}$ & $1$& $0.972$ & $6\mathrm{e}{-4}$ & $0.05$  & $0.984$ & $0.013$ & $0.041$ \\ \hline
$7$ & $2\mathrm{e}{-4}$ & $1\mathrm{e}{-12}$ & $1$ & $0$ & $0$ & $0$  & $0$ & $0$ & $0$\\ \hline\hline
\multicolumn{3}{c|}{Overall (active)}& $2.609$ & $2.337$ & $2.77\mathrm{e}{-4}$ & $0.05$  & $2.157$ & $0.011$ & $0.043$\\
\end{tabular}
\caption{BER and FER for SD-BSC$_1$ with polar coding: $A=25$, $\gamma=4.4801$, $\beta=3.5$.}
\label{tab_examplepolarcodee_SD}
\end{table}

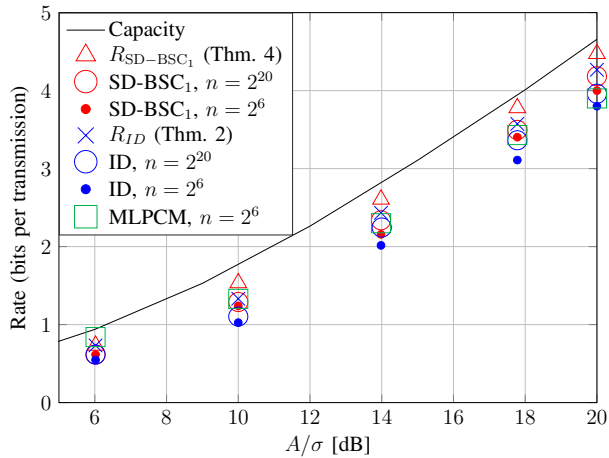
\begin{figure}[t]
    \centering
%%\begin{subfigure}[t]{0.47\textwidth}             \centering
         \tikzset{every picture/.style={scale=.85}, every node/.style={scale=.8}}
%    \input{PC_Capacity}

\begin{tikzpicture}

\begin{axis}[%
width=4.52in,
height=3.45in,
xmin=5,
xmax=20,
xlabel={$A/\sigma$ [dB]},
ymin=0,
ylabel={Rate (bits per transmission)},
ymin=0,
ymax=5,
axis background/.style={fill=white},
xmajorgrids,
ymajorgrids,
legend style={at={(axis cs: 5,5)}, anchor=north west, legend cell align=left, align=left, draw=white!15!black}
]
\addplot [color=black]
  table[row sep=crcr]{%
3	0.484399999620577\\
6	0.938500000064862\\
9	1.52889999948447\\
12	2.262600000384\\
15	3.10689999959583\\
18	4.01080000030609\\
21	4.9815000000622\\
24	5.95252679188549\\
27	6.93590799304662\\
};
\addlegendentry{Capacity}

\addplot [color=red, only marks, mark=triangle, mark options={solid, red},mark size=5pt]
  table[row sep=crcr]{%
6.0206	0.733\\
10	1.5338\\ 
13.9794	2.6057\\ 
17.78	3.778\\ 
20	4.4725\\ 
};
\addlegendentry{$R_{{\rm SD-BSC}_1}$ (Thm. \ref{theorem_SD_BSC})}

\addplot [color=red, only marks, mark=o, mark options={solid, red},mark size=5pt]
  table[row sep=crcr]{%
6.0206	0.6157\\
10	1.29\\ % FER .06
13.9794	2.3375\\ % FER .05
17.78	3.4961\\ % FER .04
20	4.1834\\ % FER .01
};
\addlegendentry{SD-BSC$_1$, $n=2^{20}$}

\addplot [color=red, only marks, mark=*, mark options={solid, red},mark size=2pt]
  table[row sep=crcr]{%
6.0206	0.6157\\
10	1.24\\ % FER .049
13.9794	2.157\\ % FER .043
17.78	3.403\\ % FER .043
20	4\\ % FER .06
};
\addlegendentry{SD-BSC$_1$, $n=2^6$}

\addplot [color=blue, only marks, mark=x, mark options={solid, blue},mark size=5pt]
  table[row sep=crcr]{%
6.0206	0.733\\
10	1.3320\\
13.9794	2.4336\\
17.78	3.5738\\
20      4.2671\\
%22.3	5.0014\\
%25.05	5.9188\\
%27	6.6097\\
};
\addlegendentry{$R_{ID}$ (Thm. \ref{theorem_independentcodingachievablerate})}

\addplot [color=blue, only marks, mark=o, mark options={solid, blue},mark size=5pt]
  table[row sep=crcr]{%
6.0206	0.6157\\ % FER <0.025 for all
10	1.1036\\
14	2.2446\\
17.78	3.3636\\
20  3.96\\
%22.3	4.74\\
%25.05	5.7284\\
%27	6.3351\\
};
\addlegendentry{ID, $n=2^{20}$}

\addplot [color=blue, only marks, mark=*, mark options={solid, blue},mark size=2pt]
  table[row sep=crcr]{%
6.0206 .5469\\ % FER .099
10 1.025\\ % FER .089
13.9794   2.0156\\ % FER .095
17.78   3.1094\\ % FER .095
20      3.8\\ % FER .075
};
\addlegendentry{ID, $n=2^6$}

\addplot [color=green!70!black, only marks, mark=square, mark options={solid, green!70!blue},mark size=5pt]
  table[row sep=crcr]{%
6.0206 .84\\ % FER .0655 OK
10 1.33\\ % FER 0.079 OK
13.9794   2.3\\ % FER .08 OK
17.78   3.43\\ % FER .067 OK
20      3.9\\ % FER .073 OK
};
\addlegendentry{MLPCM, $n=2^6$}

%% MLPCM with MSB first
%\addplot [color=green!70!black, only marks, mark=square, mark options={solid, green!70!blue},mark size=2pt]
%  table[row sep=crcr]{%
%6.0206 .3\\ % FER .07
%10 0.92\\ % FER .895
%13.9794   2.00\\ % FER .0778
%17.78   3.15\\ % FER .0834
%20      3.92\\ % FER .041
%};
%\addlegendentry{MLPCM, $n=2^6$}

\end{axis}
\end{tikzpicture}%

%    \caption{Achievable rates versus SNR at an overall FER $<0.025$.}
%     \end{subfigure}
%     \hspace{.5cm}
%     \begin{subfigure}[t]{0.47\textwidth}             \centering
%		\tikzset{every picture/.style={scale=.8}, every node/.style={scale=.8}}
%    \input{FER_VS_RATE}
%    \caption{BER/FER versus rate at SNRs $A/\sigma$ equal $14$ dB and $20$ dB, with code length $n=2^{20}$.}
%    \label{fig_fervsrate}
%     \end{subfigure}
    \caption{Performance of ID and SD-BSC$_1$ combined with polar codes, compared with $R_{\rm ID}$ (Thm. \ref{theorem_independentcodingachievablerate}), $R_{{\rm SD-BSC}_1}$ (Thm. \ref{theorem_SD_BSC}), and the achievable rate using MLPCM (all at FER $\sim 10^{-2}$).}
    \label{fig_polarcodesic}
\end{figure}

%\begin{table}[t]
%\centering
%\begin{tabular}{c"c|c|c|c|c}
%\hline
%\rowcolor{gray!60!white}
%SNR $A/\sigma$ &  $4$ & $10$ & $25$ & $60$ & $100$ \\ \hline
%\rowcolor{gray!30!white}
%Capacity 
%& $0.9419$ & $1.7584$ & $2.82$ & $3.9524$ & $4.6455$ \\\hline
%\rowcolor{gray!10!white}
%$R_{{\rm SD-BSC}_1}$ (Thm. \ref{theorem_SD_BSC}) &
%$0.7330$ & $1.5338$ & $2.6057$ & $3.7780$ & $4.4725$ \\\hline 
%\rowcolor{gray!10!white}
%SD-BSC$_1$, $n=2^{20}$ &
%$0.6157$ & $1.2900$ & $2.3375$ & $3.4961$ & $4.1834$\\\hline
%\rowcolor{gray!10!white}
%SD-BSC$_1$, $n=2^6$ & 
%$0.6157$ & $1.2400$ & $2.1570$ & $3.4030$ & $4.000$\\\hline
%\rowcolor{gray!30!white}
%$R_{ID}$ (Thm. \ref{theorem_independentcodingachievablerate}) & 
%$0.7330$ & $1.3320$ & $2.4336$ & $3.5738$ & $4.2671$ \\\hline
%\rowcolor{gray!30!white}
%ID, $n=2^{20}$ & 
%$0.6157$ & $1.1036$ & $2.2446$ & $3.3636$ & $3.9600$\\\hline
%\rowcolor{gray!30!white}
%ID, $n=2^6$ &
%$0.5469$ & $1.0250$ & $2.0156$ & $3.1094$ & $3.8000$\\\hline
%\rowcolor{gray!10!white}
%MLPCM, $n=2^6$ & 
%$0.8400$ & $1.3300$ & $2.3000$ & $3.4300$ & $3.9000$\\\hline
%\end{tabular}
%\caption{Performance of ID and SD-BSC$_1$ combined with polar codes, compared with $R_{\rm ID}$ (Thm. \ref{theorem_independentcodingachievablerate}), $R_{{\rm SD-BSC}_1}$ (Thm. \ref{theorem_SD_BSC}), and the achievable rate using MLPCM (all at FER $\sim 10^{-2}$). The channel capacity is also given for reference.}
%\label{fig_polarcodesic}
%\end{table}

Fig. \ref{fig_polarcodesic} shows the achievable rates using the ID and SD-BSC$_1$ schemes implemented using polar codes for a range of SNRs. The figure shows the achievability of rates close to $R_{\rm ID}$ (Thm. \ref{theorem_independentcodingachievablerate}) and $R_{{\rm BSD-BSC}_1}$ (Thm \ref{theorem_SD_BSC}), where the gap to $R_{\rm ID}$ and $R_{{\rm BSD-BSC}_1}$ is due to the use of finite $n$ and the used polar coding packages. The gap can be reduced using stronger BCC, or using shaping which may be helpful to improve performance when using short codes as in \cite{tw_C}. To compare with standard MLC/MSD, we also include the performance of multi-level polar-coded modulation (MLPCM) with set partitioning. The proposed schemes achieve rates which are comparable with MLPCM while only using BCC. The advantage compared to MLPCM is that our proposed schemes use binary operations, whereas MLPCM requires real-valued operations, which can impact the transmitter and receiver processing and memory requirements in practice.
 
%Fig \ref{fig_fervsrate} shows the overall FER and BER of the ID  scheme implemented using polar codes at different rates for SNRs $A/\sigma$ equal to $14$ and $20$ dB. For an SNR of $14$ dB, $\beta=5$ and $\gamma=4.4801$ are selected, whereas $\beta=4.5$ and $\gamma=4.9601$ at an SNR of $20$ dB (numerically optimized). Note that the BER and FER decrease sharply when the rate drops below $R_{\rm ID}-\Delta$ where $R_{\rm ID}$ is the theoretical achievable rate given in Theorem \ref{theorem_independentcodingachievablerate}, and where $\Delta\approx 0.3$ bits. 
%Decreasing the overall BER and FER can achieved by decreasing the rate for the worst bit-pipe first, then the second worst, and so on, noting that bit-pipes with low capacities and high BERs/FERs increase the overall BER/FER but do not carry high rates. During this process, the BER/FER decreases rapidly for a slight decrease in the rate. 

To put these results in context with the literature, we note that \cite{prob_4pam}, \cite{tw_C}, and \cite{tw_A} achieve rates of $1$, $1.25$, and $2$ bit per transmission at an FER of $10^{-2}$ and an SNR of $\frac{A}{\sigma}\approx 7.5$, $8.3$, and $11.5$ dB, respectively. The SD-BSC$_1$ scheme achieves a rate of $0.89$, $0.94$, and $1.63$ bits per transmission at $n=64$ (theoretically $0.95$, $1.01$, and $1.86$ as $n\to\infty$) at these SNRs, respectively. The advantage of the proposed schemes (such as the SD-BSC scheme) is that they are general (can produce any PAM order), easily applicable for the IM/DD channel with a peak constraint only or a peak and average constraint, and can be implemented in a straightforward way using BCC for decoding from the binary channel output instead of the soft channel output. Also, although our scheme is generally described for $N$ bit-pipes, only $1$-$3$ bit-pipes will be active in the examples above leading to comparable complexity as \cite{prob_4pam,tw_A}.

\section{Conclusion}
\label{sec_conclusion}
We proposed the VBC as an approximation of the IM/DD channel (an AWGN with positive input, peak and average constraints). The VBC is a collection of $N$ bit-pipes, representing a channel with a binary vector input, noise, and output. This decomposition is capacity-preserving as it carefully considers interactions between bits on different bit-pipes at the receiver. In addition, we presented five coding schemes for the VBC model intending to (i) approach capacity, and (ii) design a practical coding scheme for the IM/DD channel which can be implemented using codes for binary-input binary-output channels. The achievable rates of the schemes are evaluated and compared to capacity and capacity bounds. The achievable rates of the \textit{state-assisted decoding} and its simplification, which uses the concept of a channel with state, are nearly equal to capacity at moderate and high SNR as shown in Fig. \ref{capfigAllSchemes}. The \textit{independent decoding} scheme, which treats each of the VBC's bit-pipes independently, secures good achievable rates (about $0.3$ bits away from capacity at moderate and high SNR) while being a practical coding scheme for the VBC that is realizable using polar codes. At low SNR, the  \textit{carry-over-assisted decoding} scheme and its simplification have the best performance in terms of achievable rates, closing the gap to capacity that other schemes exhibit at low SNR. 

This work can be extended in several ways. One way is to design coding schemes for a multi-user setting, i.e., broadcast and multiple access channels. Another extension would be to consider different channel models, such as a power-constrained Gaussian channel. Also, the development of similar schemes for the IM/DD channel different channel codes, a combination of the SD and the CD schemes, or with oversampling (as in \cite{tw_D}) are all interesting research directions.

\bibliographystyle{IEEEtran}
%\bibliography{../../../tex/myBib}
%\bibliography{bibliography}

\begin{thebibliography}{10}
\providecommand{\url}[1]{#1}
\csname url@samestyle\endcsname
\providecommand{\newblock}{\relax}
\providecommand{\bibinfo}[2]{#2}
\providecommand{\BIBentrySTDinterwordspacing}{\spaceskip=0pt\relax}
\providecommand{\BIBentryALTinterwordstretchfactor}{4}
\providecommand{\BIBentryALTinterwordspacing}{\spaceskip=\fontdimen2\font plus
\BIBentryALTinterwordstretchfactor\fontdimen3\font minus
  \fontdimen4\font\relax}
\providecommand{\BIBforeignlanguage}[2]{{%
\expandafter\ifx\csname l@#1\endcsname\relax
\typeout{** WARNING: IEEEtran.bst: No hyphenation pattern has been}%
\typeout{** loaded for the language `#1'. Using the pattern for}%
\typeout{** the default language instead.}%
\else
\language=\csname l@#1\endcsname
\fi
#2}}
\providecommand{\BIBdecl}{\relax}
\BIBdecl

\bibitem{ourpaper}
S.~Bahanshal and A.~Chaaban, ``{A binary decomposition and transmission schemes
  for the peak-constrained IM/DD channel},'' in \emph{IEEE Int. Symp. Inf.
  Theory (ISIT)}, Melbourne, Australia, July 2021, pp. 3103--3108.

\bibitem{hew_opticalwirelesscomp}
H.~Haas, J.~Elmirghani, and I.~White, ``{Optical wireless communication},''
  \emph{Phil. Trans. Series A, Math., Phy., Eng. Sci.}, 2020.

\bibitem{chaaban2020tutorial}
A.~Chaaban, Z.~Rezki, and M.-S. Alouini, ``{On the capacity of
  intensity-modulation direct-detection Gaussian optical wireless communication
  channels: A tutorial},'' \emph{IEEE Commun. Surveys Tuts.}, vol.~24, no.~1,
  pp. 455--491, 2021.

\bibitem{ku_fsosurvey}
M.-A. Khalighi and M.~Uysal, ``{Survey on free space optical communication: A
  communication theory perspective},'' \emph{IEEE Commun. Surveys Tuts.},
  vol.~16, no.~4, pp. 2231--2258, 2014.

\bibitem{hwnc_opticalstory}
D.~J.~T. {Heatley}, D.~R. {Wisely}, I.~{Neild}, and P.~{Cochrane}, ``Optical
  wireless: The story so far,'' \emph{IEEE Commun. Mag.}, vol.~36, no.~12, pp.
  72--74, 1998.

\bibitem{ch_ptwiresur}
A.~Chaaban and S.~Hranilovic, ``Capacity of optical wireless communication
  channels,'' \emph{Phil. Trans. Series A. Math., Phy., Eng. Sci.}, vol. 378,
  p. 20190184, Mar. 2020.

\bibitem{AdvancedOpticalWireless}
S.~Arnon, J.~Barry, G.~Karagiannidis, R.~Schober, and M.~Uysal, \emph{{Advanced
  optical wireless communication systems}}.\hskip 1em plus 0.5em minus
  0.4em\relax Cambridge University Press, 2012.

\bibitem{book_capacityofopticalchannels}
K.-P. Ho, \emph{Phase-modulated optical communication systems}.\hskip 1em plus
  0.5em minus 0.4em\relax Springer, 2005.

\bibitem{WIinfraredCom}
J.~Kahn and J.~Barry, ``Wireless infrared communications,'' \emph{Proc. IEEE},
  vol.~85, no.~2, pp. 265--298, 1997.

\bibitem{lmw_bounds}
A.~{Lapidoth}, S.~M. {Moser}, and M.~A. {Wigger}, ``On the capacity of
  free-space optical intensity channels,'' \emph{IEEE Trans. Inf. Theory},
  vol.~55, no.~10, pp. 4449--4461, 2009.

\bibitem{fh_signalling}
A.~A. {Farid} and S.~{Hranilovic}, ``Channel capacity and non-uniform
  signalling for free-space optical intensity channels,'' \emph{IEEE J. Sel.
  Areas in Commun.}, vol.~27, no.~9, pp. 1553--1563, 2009.

\bibitem{FaridHranilovic}
A.~A. Farid and S.~Hranilovic, ``{Capacity bounds for wireless optical
  intensity channels with Gaussian noise},'' \emph{IEEE Trans. Inf. Theory},
  vol.~56, no.~12, pp. 6066--6077, Dec. 2010.

\bibitem{ChaabanMorvanAlouini}
A.~Chaaban, J.-M. Morvan, and M.-S. Alouini, ``{Free-space optical
  communications: Capacity bounds, approximations, and a new sphere-packing
  perspective},'' \emph{IEEE Trans. Commun.}, vol.~64, no.~3, pp. 1176--1191,
  Mar. 2016.

\bibitem{bounds_mckellips}
A.~McKellips, ``Simple tight bounds on capacity for the peak-limited
  discrete-time channel,'' in \emph{IEEE Int. Symp. Inf. Theory (ISIT)},
  Chicago, USA, June 2004, pp. 348--348.

\bibitem{bounds_Thangaraj}
A.~Thangaraj, G.~Kramer, and G.~B\"{o}cherer, ``{Capacity bounds for
  discrete-time, amplitude-constrained, additive white Gaussian noise
  channels},'' \emph{IEEE Trans. Inf. Theory}, vol.~63, no.~7, pp. 4172--4182,
  2017.

\bibitem{bounds_clerckx}
B.~Rassouli and B.~Clerckx, ``An upper bound for the capacity of
  amplitude-constrained scalar awgn channel,'' \emph{IEEE Commun. Lett.},
  vol.~20, no.~10, pp. 1924--1926, 2016.

\bibitem{smith_finitemasspoints}
J.~G. Smith, ``{The information capacity of amplitude- and variance-constrained
  sclar Gaussian channels},'' \emph{Inf. Control}, vol.~18, no.~3, pp.
  203--219, 1971.

\bibitem{chk_discreteinputdist}
T.~H. {Chan}, S.~{Hranilovic}, and F.~R. {Kschischang}, ``{Capacity-achieving
  probability measure for conditionally Gaussian channels with bounded
  inputs},'' \emph{IEEE Trans. Inf. Theory}, vol.~51, no.~6, pp. 2073--2088,
  2005.

\bibitem{ldpc_gallager}
R.~Gallager, ``Low-density parity-check codes,'' \emph{IRE Trans. Inf. Theory},
  vol.~8, no.~1, pp. 21--28, 1962.

\bibitem{arikan_channelpolarization}
E.~Ar{\i}kan, ``Channel polarization: A method for constructing
  capacity-achieving codes for symmetric binary-input memoryless channels,''
  \emph{IEEE Trans. Inf. Theory}, vol.~55, no.~7, pp. 3051--3073, 2009.

\bibitem{erdal_ontheoriginofpolarcodes}
------, ``On the origin of polar coding,'' \emph{IEEE J. Sel. Areas Commun.},
  vol.~34, no.~2, pp. 209--223, 2015.

\bibitem{wfh_mlc}
U.~{Wachsmann}, R.~F.~H. {Fischer}, and J.~B. {Huber}, ``Multilevel codes:
  Theoretical concepts and practical design rules,'' \emph{IEEE Trans. Inf.
  Theory}, vol.~45, no.~5, pp. 1361--1391, 1999.

\bibitem{adt_deterministic}
A.~S. {Avestimehr}, S.~N. {Diggavi}, and D.~N.~C. {Tse}, ``Wireless network
  information flow: A deterministic approach,'' \emph{IEEE Trans. Inf. Theory},
  vol.~57, no.~4, pp. 1872--1905, 2011.

\bibitem{BICM}
E.~Zehavi, ``{8-PSK trellis codes for a Rayleigh channel},'' \emph{IEEE Trans.
  Commun.}, vol.~40, no.~5, pp. 873--884, 1992.

\bibitem{multilevel_nonuni}
J.~Jiang and K.~R. Narayanan, ``{Multilevel coding for channels with
  non-uniform inputs and rateless transmission over the BSC},'' in \emph{IEEE
  Int. Symp. Inf. Theory (ISIT)}, Seatle, USA, July 2006, pp. 518--522.

\bibitem{book_gallager1968information}
R.~Gallager, \emph{{Information theory and reliable communication}}.\hskip 1em
  plus 0.5em minus 0.4em\relax Wiley, 1968.

\bibitem{tw_C}
C.~Runge, T.~Wiegart, D.~Lentner, and T.~Prinz, ``Multilevel binary polar-coded
  modulation achieving the capacity of asymmetric channels,'' in \emph{IEEE
  Int. Symp. Inf. Theory (ISIT)}, Espoo, Finland, June 2022.

\bibitem{newmultilevel_ih}
H.~Imai and S.~Hirakawa, ``A new multilevel coding method using
  error-correcting codes,'' \emph{IEEE Trans. Inf. Theory}, vol.~23, no.~3, pp.
  371--377, 1977.

\bibitem{c_power_bwefficient_ldpc}
P.~Limpaphayom and K.~Winick, ``{Power- and bandwidth-efficient communications
  using LDPC codes},'' \emph{IEEE Trans. Commun.}, vol.~52, no.~3, pp.
  350--354, 2004.

\bibitem{d_performance_ldpc}
Y.~Kofman, E.~Zehavi, and S.~Shamai, ``Performance analysis of a multilevel
  coded modulation system,'' \emph{IEEE Trans. Commun.}, vol.~42, pp. 299--312,
  03 1994.

\bibitem{st_feedbackcap}
C.~Suh and D.~N.~C. Tse, ``{Feedback capacity of the Gaussian interference
  channel to within 2 bits},'' \emph{IEEE Trans. Inf. Theory}, vol.~57, no.~5,
  pp. 2667--2685, 2011.

\bibitem{ht_interferencemitigation}
I.-H. Wang and D.~N.~C. Tse, ``Interference mitigation through limited receiver
  cooperation,'' \emph{IEEE Trans. Inf. Theory}, vol.~57, no.~5, pp.
  2913--2940, 2011.

\bibitem{CIT-042}
\BIBentryALTinterwordspacing
A.~S. Avestimehr, S.~N. Diggavi, C.~Tian, and D.~N.~C. Tse, ``An approximation
  approach to network information theory,'' \emph{Foundations Trends Commun.
  Inf. Theory}, vol.~12, no. 1-2, pp. 1--183, 2015. [Online]. Available:
  \url{http://dx.doi.org/10.1561/0100000042}
\BIBentrySTDinterwordspacing

\bibitem{ElgamalKim}
A.~E. Gamal and Y.-H. Kim, \emph{{Network information theory}}.\hskip 1em plus
  0.5em minus 0.4em\relax Cambridge University Press, 2011.

\bibitem{tw_A}
T.~Prinz, T.~Wiegart, D.~Plabst, S.~Calabr\`{o}, G.~B\"{o}cherer,
  N.~Stojanovi\'{c}, and T.~Rahman, ``{PAM-6 coded modulation for IM/DD
  channels with a peak-power constraint},'' in \emph{Int. Symp. Topics Coding
  (ISTC)}, Montreal, Canada, Aug. 2021.

\bibitem{prob_4pam}
T.~Wiegart, F.~Da~Ros, M.~P. Yankov, F.~Steiner, S.~Gaiarin, and R.~D. Wesel,
  ``{Probabilistically shaped 4-PAM for short-reach IM/DD links with a peak
  power constraint},'' \emph{J. Lightw. Technol.}, vol.~39, no.~2, pp.
  400--405, 2021.

\bibitem{tw_B}
T.~Prinz, T.~Wiegart, D.~Plabst, T.~Rahman, M.~S. Hossain, N.~Stojanovi\'{c},
  S.~Calabr\`{o}, N.~Hanik, and G.~Kramer, ``{Comparison of PAM-6 modulations
  for short-reach fiber-optic links with intensity modulation and direct
  detection},'' in \emph{https://arxiv.org/pdf/2205.05460.pdf}, 2022.

\bibitem{ILX_LD}
\emph{{An overview of laser diode characteristics - AN05}}, ILX Lightwave
  Corporation, 2005.

\bibitem{d_poissonchannel}
A.~D. Wyner, ``{Capacity and error exponent for the direct detection photon
  channel--Part I},'' \emph{IEEE Trans. Inf. Theory}, vol.~34, no.~6, pp.
  1449--1461, Nov. 1988.

\bibitem{Moser}
S.~M. Moser, ``{Capacity results of an optical intensity channel with
  input-dependent Gaussian noise},'' \emph{IEEE Trans. Inf. Theory}, vol.~58,
  no.~1, pp. 207--223, Jan. 2012.

\bibitem{Safari_DependentNoise}
M.~Safari, ``Efficient optical wireless communication in the presence of
  signal-dependent noise,'' in \emph{IEEE Int. Conf. Commun. Workshop (ICCW)},
  London, UK, June 2015, pp. 1387--1391.

\bibitem{cover_elementsIT2}
T.~M. Cover and J.~A. Thomas, \emph{Elements of information theory},
  2nd~ed.\hskip 1em plus 0.5em minus 0.4em\relax New York: Wiley, 2006.

\bibitem{vw_channelwitherasure}
S.~{Verd\'u} and T.~{Weissman}, ``The information lost in erasures,''
  \emph{IEEE Trans. Inf. Theory}, vol.~54, no.~11, pp. 5030--5058, 2008.

\bibitem{NagarjunaSiegel}
K.~Nagarjuna and P.~H. Siegel, ``Universal polar coding for asymmetric
  channels,'' in \emph{IEEE Inf. Theory Workshop (ITW)}, Guangzhou, China, Nov.
  2018, pp. 1--5.

\bibitem{HondaYamamoto}
J.~Honda and H.~Yamamoto, ``Polar coding without alphabet extension for
  asymmetric models,'' \emph{IEEE Trans. Inf. Theory}, vol.~59, no.~12, pp.
  7829--7838, 2013.

\bibitem{Polyanskiy}
Y.~Polyanskiy, H.~V. Poor, and S.~Verdu, ``Channel coding rate in the finite
  blocklength regime,'' \emph{IEEE Transactions on Information Theory},
  vol.~56, no.~5, pp. 2307--2359, 2010.

\bibitem{vangala2016practical}
\BIBentryALTinterwordspacing
H.~Vangala, Y.~Hong, and E.~Viterbo. (2016) Polar coding algorithms in matlab.
  Accessed May 31, 2022. [Online]. Available:
  \url{https://ecse.monash.edu/staff/eviterbo/polarcodes.html}
\BIBentrySTDinterwordspacing

\bibitem{PolarCodeGitHub}
A.~R. AlHamali, ``{A Matlab implementation of polar codes},''
  \url{https://github.com/AbdulRahmanAlHamali/polar-codes-matlab}, accessed:
  2022-12-04.

\bibitem{tw_D}
D.~Plabst, T.~Prinz, T.~Wiegart, T.~Rahman, N.~Stojanovi\'{c}, S.~Calabr\`{o},
  N.~Hanik, and G.~Kramer, ``Achievable rates for short-reach fiber-optic
  channels with direct detection,'' \emph{J. Lightw. Tech.}, vol.~40, no.~12,
  pp. 3602--3613, 2022.

\end{thebibliography}


\end{document}